\documentclass[final,3p,times,twocolumn]{elsarticle}

\usepackage{amssymb}
\usepackage{amsmath}

\usepackage[utf8]{inputenc}
\usepackage{bm}
\usepackage{booktabs}
\usepackage{color,soul}
\usepackage{cuted}
\usepackage{float}
\usepackage[T1]{fontenc}
\usepackage{makecell}
\usepackage{multirow}
\usepackage{pgffor}
\usepackage{placeins}
\usepackage{stfloats}
\usepackage{subcaption}
\usepackage{tabularx}
\usepackage{textcomp, gensymb}
\usepackage[table]{xcolor}

\usepackage{hyperref}
\usepackage{cleveref}

\usepackage{tikz}
\usetikzlibrary{shapes.geometric, arrows, positioning, calc, fit}
\usepackage{tikz-3dplot}
\usepackage{pgfplots}
\pgfplotsset{compat=1.18}
\usepackage{xcolor}
\usepackage{pgfplotstable}
\usepgfplotslibrary{statistics}

\newcommand{\githubrepo}{\url{http://github.com/ICterraOfficial/midas} (Retrieved on 14 August 2025)}

\journal{ }

\begin{document}

\begin{frontmatter}

\title{DoSReMC: Domain Shift Resilient Mammography Classification using Batch Normalization Adaptation\tnoteref{t1}}
\tnotetext[t1]{This work was conducted under the auspices of ICterra Information and Communication Technologies and supported by the company’s research and development initiative.}

\author[1]{\texorpdfstring{Uğurcan Akyüz \corref{cor1}}}
\ead{ugurcan.akyuz0@gmail.com}
\cortext[cor1]{Corresponding author}

\author[1]{Deniz Katircioglu-Öztürk}
\ead{denizkatircioglu@gmail.com}

\author[1]{Emre K. Süslü}
\author[1]{Burhan Keleş}
\author[1,3]{Mete C. Kaya}
\author[2]{Gamze Durhan}
\author[2]{Meltem G. Akpınar}
\author[2]{Figen B. Demirkazık}
\author[3]{Gözde B. Akar}

\affiliation[1]{organization={ICterra Information and Communication Technologies},
            city={Çankaya, Ankara},
            postcode={06531}, 
            country={Türkiye}}

\affiliation[2]{organization={Department of Radiology, Hacettepe University Faculty of Medicine},
            city={Altındağ, Ankara},
            postcode={06230}, 
            country={Türkiye}}

\affiliation[3]{organization={Department of Electrical and Electronics Engineering, Middle East Technical University},
            city={Çankaya, Ankara},
            postcode={06800}, 
            country={Türkiye}}

\begin{abstract} 
Numerous deep learning–based solutions have been developed for the automatic recognition of breast cancer using mammography images. However, their performance often declines when applied to data from different domains, primarily due to domain shift — the variation in data distributions between source and target domains. This performance drop limits the safe and equitable deployment of AI in real-world clinical settings. In this study, we present DoSReMC (Domain Shift Resilient Mammography Classification), a batch normalization (BN) adaptation framework designed to enhance cross-domain generalization without retraining the entire model. Using three large-scale full-field digital mammography (FFDM) datasets — including HCTP, a newly introduced, pathologically confirmed in-house dataset — we conduct a systematic cross-domain evaluation with convolutional neural networks (CNNs). Our results demonstrate that BN layers are a primary source of domain dependence: they perform effectively when training and testing occur within the same domain, and they significantly impair model generalization under domain shift. DoSReMC addresses this limitation by fine-tuning only the BN and fully connected (FC) layers, while preserving pretrained convolutional filters. We further integrate this targeted adaptation with an adversarial training scheme, yielding additional improvements in cross-domain generalizability while reducing the computational cost of model training. DoSReMC can be readily incorporated into existing AI pipelines and applied across diverse clinical environments, providing a practical pathway toward more robust and generalizable mammography classification systems.
\end{abstract}


\end{frontmatter}


\section{Introduction}
Breast cancer is among the most prevalent forms of cancer worldwide, with 2.3 million new cases reported in 2022, making it the second most common type of cancer globally \citep{bray2021global}. In 2024, it was projected to become the most common cancer among women in the United States, with an estimated 310,720 new cases and 42,250 deaths, making it the second leading cause of cancer-related mortality \citep{siegel2024cancer}. Early detection through mammography has been shown to significantly improve survival rates, with a five-year survival rate exceeding 90\% when the disease is diagnosed in an early, localized stage. As a vital tool in the early detection of breast cancer, screening mammography is the most common diagnostic technique used to identify abnormalities such as calcification, mass, and structural distortion of the breast \citep{ren2022global}. It plays a crucial role in improving the chances of successful treatment and reducing mortality rates. Studies show that regular mammography screening can reduce breast cancer mortality by approximately 20–40\% among women aged 50–69 \citep{katalinic2020breast}.

Mammography is an imaging procedure that utilizes X-ray technology to generate high-resolution images of breast tissue. The process involves a specialized device that generates two-dimensional images of the breast from multiple angles, predominantly the craniocaudal (CC) and mediolateral oblique (MLO) views. During this procedure, a compression paddle is used to gently compress the breast, which helps spread the tissue for a clearer image, reduces motion blurring, and minimizes the amount of radiation required. X-rays passing through the breast tissue are subsequently captured by digital detectors, which measure their intensity and convert that data into digital signals to generate mammograms. 

Radiologists then interpret mammograms using the Breast Imaging Reporting and Data System (BI-RADS), which standardizes reporting and aids clinical decision-making. The scale ranges from 0 (incomplete) to 6 (biopsy-proven malignancy), with intermediate scores reflecting increasing suspicion of cancer. Lower scores indicate normal or benign findings, while higher scores typically lead to recommendations for short-term follow-up or biopsy. This standardized framework improves consistency in diagnosis and helps guide appropriate patient management.

Despite its clinical utility, interpreting mammography images is challenging. Malignant findings can occasionally be overlooked, particularly when obscured by dense breast tissue or when they are too small relative to the resolution of the image. This can lead to an increased rate of false negatives. While some studies have shown that mammography screening has significantly reduced breast cancer-related deaths \citep{duffy2002impact, schopper2009effective, kalager2010effect, katalinic2020breast}, it still results in a relatively high number of false positives where mammogram examination suggests the presence of cancer or an abnormality that, upon further testing, turns out to be benign\citep{brodersen2013long, nelson2016factors}. 

In order to reduce false positive and false negative rates and ease the workload of radiologists, numerous artificial intelligence systems have been developed. Contemporary research on artificial intelligence is mainly based on CNNs\citep{zhang2018classification, tan2017breast, lu2019classification}. Yet clinical deployment of the CNNs is hindered by domain shift, which typically stems from differences in pixel intensity distributions and acquisition settings across imaging devices. Addressing this robustness gap is essential for safe and equitable AI-assisted breast-cancer screening.

In this work, we introduce \textit{DoSReMC}, a BN adaptation framework that improves cross-domain generalization. The main contributions of this research are as follows:

\begin{itemize}
  \item We introduce a new in-house mammography dataset, the Hacettepe-Mammo Dataset (HCTP), which comprises 157,463 high-resolution full-field digital mammography images with findings pathologically confirmed. This is, to our knowledge, the largest mammography dataset created in Türkiye, including radiological reports and pathologically proven diagnoses. The dataset also encompasses a broad spectrum of radiological findings, including calcifications (ductal, punctate), masses (spiculated, lobular, indistinct, and circumscribed), asymmetric distortions, and lymphadenopathies (LAP). 

  \item We conduct the first comprehensive analysis of domain shift in mammography classification from an architectural perspective, showing how scanner-dependent pixel intensity distributions affect BN layers and, in turn, degrade cross-domain generalization.
  
  \item We demonstrate that tuning only the BN layers and FC layers can achieve comparable results to fine-tuning the entire model.
  
  \item We propose a partial domain-adversarial training (DAT) strategy that fine-tunes only the BN and FC layers, freezing convolutional layers, to address domain shift in mammography classification. This takes a different path from the conventional approaches that adapt the entire network, enabling more targeted and cost-efficient adaptation while preserving pretrained convolutional representations.
\end{itemize}

\section{Related Work}
\subsection{Deep Learning Approaches in Mammography}
High-resolution and multi-view inputs are crucial for accurate mammography analysis. While many CNN-based approaches downscale images to reduce computational cost, this can obscure small but clinically significant structures. To investigate the impact of resolution, a deep convolutional network \citep{geras2017high} was designed to handle high-resolution mammography images. Additionally, as this research indicates, mammography images from both views (CC and MLO) can provide crucial information regarding malignancies. Several studies have likewise utilized multi-view mammography images for assessment \citep{khan2019multi,wang2018breast,kyono2018mammo,lopez2022hypercomplex, pathak2025breast}.
 
Deep learning has been applied to a wide range of mammography tasks. For classification and localization, \citep{ertosun2015probabilistic, fathy2019deep, barnett2021case, zhou2019weakly} proposed CNN-based models for mammography and magnetic resonance (MR) images. Detection methods have been developed for specific suspicious findings, with some studies targeting masses \citep{agarwal2019automatic, cao2021breast, su2022yolo, zheng2023deep} and others focusing on calcifications \citep{yurdusev2023detection, wang2019automated, sakaida2023development}. Beyond detection of suspicious findings, \citep{yala2021toward, yala2022multi} proposed a novel deep learning–based breast cancer risk prediction model, evaluating it across multiple datasets, including the CSAW-CC dataset \citep{dembrower2020multi}. Their results highlight the potential of mammography-based AI systems for robust long-term risk assessment.

\subsection{Weakly Supervised and Label-Efficient Methods}
While these studies demonstrate the significant potential of deep learning in assisting with the analysis of mammograms, the process of collecting and labeling mammogram images continues to be a challenging task. The manual annotation of findings is extremely time-consuming and labor-intensive. To overcome this, \citep{liu2021weakly, tang2021leveraging, bakalo2019classification} leveraged a weakly supervised learning strategy to train deep models. The results showed that weakly supervised methods can mitigate the stringent labeling requirements. \citep{shen2021interpretable} proposed a novel model that utilizes a weakly supervised approach to classify mammography images and highlight suspicious regions. The model is both memory-efficient and capable of achieving radiologist-level accuracy. However, its architecture is intricate, and more specifically, the saliency maps generated by the global module may diminish the contributions of the local module.

\subsection{Domain Shift and Adaptation}
Despite significant advancements, deploying these models in real-world clinical settings remains challenging due to domain shift. Deep learning models require substantial and diverse datasets to attain high sensitivity and generalizability. However, in the healthcare domain, datasets that are collected from different hospitals, clinics, or publicly available repositories often exhibit variations in image acquisition techniques, scanner manufacturers, and annotation protocols \citep{jimenez2023memory, roth2020federated, perone2019unsupervised, karani2018lifelong}. In the field of mammography, these disparities stem from variations in X-ray parameters (e.g., kilovoltage peak, exposure time, and anode/filter combination), detector resolution, bit depth, and hardware configurations, all of which influence pixel intensity distributions and image quality. Such disparities alter the statistical properties of images, causing models trained in one domain to perform poorly when applied to others.

To mitigate this challenge, several studies have investigated the underlying causes and proposed strategies for domain adaptation \citep{guan2021domain, karani2018lifelong}. For example, \citep{ghafoorian2017transfer} explored transfer learning techniques for brain lesion segmentation, while \citep{gu2019progressive} introduced a multi-step domain adaptation approach for skin cancer classification, which improves model generalization across different imaging sources. Similarly, in the field of endoscopy, \citep{laiz2019using} utilized the triplet loss function to enhance model performance, enabling a model trained on images from older capsule endoscopy devices to adapt effectively to images from newer devices.

\subsection{Domain Shift in Mammography}
Recent studies have addressed the domain shift in breast cancer detection and classification using various approaches. \citep{garrucho2022domain} performed an extensive analysis revealing that transformer-based detection models offer improved robustness to domain shift; however, their work focused solely on mass detection and did not consider calcifications, which are critical mammographic findings. Transfer learning strategies have been explored to enhance cross-dataset breast cancer classification \citep{kumar2017cross}, while adversarial learning techniques have been proposed to mitigate domain-induced performance degradation and improve breast tissue segmentation \citep{ryan2021unsupervised}.

To develop domain-invariant models, \citep{quintana2024contrastive} utilized supervised contrastive learning combined with a two-step training strategy, demonstrating improved generalization across diverse datasets. Similarly, \citep{yala2021toward} applied a conditional adversarial training scheme to build a device-invariant breast cancer risk prediction model; however, their method was not evaluated on images from multiple device vendors, limiting its validation across real-world scanner variability. Additionally, \citep{lauritzen2023robust} introduced a data augmentation approach designed to enhance mammography model adaptability across scanner vendors, which led to improved breast cancer risk prediction.

Despite these promising advances, limited research has investigated domain shift in mammography from an architectural perspective, particularly the influence of BN layers—an aspect that forms the core of our study.

\section{Datasets}
Breast cancer is a major focus in machine learning research, and numerous open-source and private mammography datasets have been developed to support advancements in this area. However, many of these datasets suffer from key limitations: some offer only a limited number of samples, many are not publicly available, and most lack representation from diverse geographic regions — particularly underrepresented populations.

To bridge this gap, we introduce a new large-scale mammography dataset, the Hacettepe-Mammo (HCTP) dataset, constructed using patient data collected in Türkiye. HCTP not only offers a high volume of pathologically confirmed cases but also captures region-specific radiological characteristics that are underrepresented in existing datasets.

In addition to HCTP, we incorporate two widely used public datasets — VinDr-Mammo \citep{nguyen2023vindr} and CSAW-CC \citep{Strand2022CSAW-CC} — to analyze cross-dataset generalization and address domain shift. For clarity, these datasets are referred to as HCTP, VinDr, and CSAW throughout this paper.

\subsection{HCTP Dataset}
\subsubsection{Data Collection}
\definecolor{frenchblue}{rgb}{0.0, 0.45, 0.73}

\begin{figure}[b]
    \centering
    \begin{tikzpicture}
        \begin{axis}[
            ybar,
            symbolic x coords={34-46, 46-57, 57-68, 68-79, 79-90},
            xtick=data,
            ymin=0,
            ymax=22500,
            xlabel={Patient Age},
            ylabel={Number of Studies},
            ylabel near ticks,
            xticklabel style={font=\small},
            width=0.45\textwidth,
            height=5.5cm,
            bar width=18pt,
            nodes near coords,
            every node near coord/.append style={font=\footnotesize, color=black},
            ytick=\empty,
            xtick pos=bottom,
        ]
        \addplot coordinates {
            (34-46, 10044) (46-57, 18963) (57-68, 12640) (68-79, 4261) (79-90, 501)
        } 
        [fill=frenchblue];
        \end{axis}
    \end{tikzpicture}
    \caption{Distribution of Patient Age per study in the HCTP dataset.}
    \label{dataset:HCTPage_dist}
\end{figure}
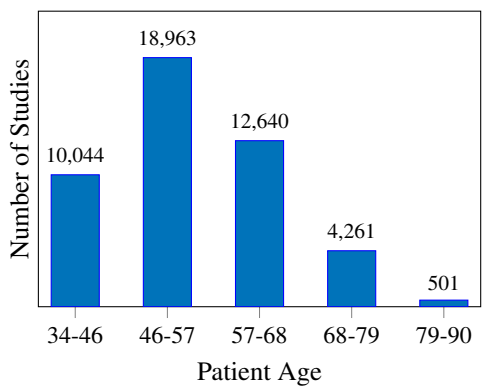

The HCTP dataset was created in collaboration with the Department of Radiology at the Faculty of Medicine, Hacettepe University, one of the leading university hospitals in Türkiye. Anonymized patient data (between 2009 and 2020), including mammography images, radiology, and pathology reports, were acquired from the digital archives of the hospital information system and the picture archiving and communication system (PACS).

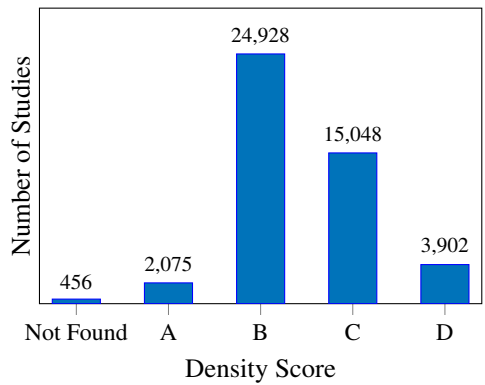
\begin{figure}[t]
    \centering
    \begin{tikzpicture}
        \begin{axis}[
            ybar,
            symbolic x coords={Not Found, A, B, C, D},
            xtick=data,
            ymin=0,
            ymax=29500,
            xlabel={Density Score},
            ylabel={Number of Studies},
            ylabel near ticks,
            xticklabel style={font=\small},
            width=0.45\textwidth,
            height=5.5cm,
            bar width=18pt,
            nodes near coords,
            every node near coord/.append style={font=\footnotesize, color=black},
            ytick=\empty,
            xtick pos=bottom,
        ]
        \addplot coordinates {(Not Found, 456) (A, 2075) (B, 24928) (C, 15048) (D, 3902)} 
        [fill=frenchblue];
        \end{axis}
    \end{tikzpicture}
    \caption{Distribution of Density scores per study in the HCTP dataset.}
    \label{dataset:HCTPdense_dist}
\end{figure}

The dataset comprises 46,409 studies from 22,279 unique patients between 35 and 90 years of age (see the \autoref{dataset:HCTPage_dist}) and includes a total of 157,463 full-field digital mammography images in monochrome-2 format. By the term 'study', we refer to the mammography screening of a patient on a specific date. The dataset comprises images with two distinct resolutions: 2394x3062 and 2294x1914, all captured using the GE Senographe Essential mammography device. Mammography images include various malignant and benign findings, including cases of masses, ductal carcinoma, lobular carcinoma, spiculated lesions, and lymphadenopathy (LAP). The dataset also encompasses various breast density levels (see the \autoref{dataset:HCTPdense_dist}).Furthermore, it comprises patients who are under follow-up, and the distribution of their visit frequency is illustrated in \autoref{dataset:HCTPvisit_dist_hist}.

\begin{figure}[t]
    \centering
    \begin{tikzpicture}
        \begin{axis}[
            ybar,
            symbolic x coords={1, 2, 3, 4, 5, 6, 7, 8, 9},
            xtick=data,
            ymin=0,
            ymax=15000,
            xlabel={Number of visits},
            ylabel={Number of patients},
            ylabel near ticks,
            xticklabel style={font=\small},
            width=0.45\textwidth,
            height=5.5cm,
            bar width=15.5pt,
            nodes near coords,
            every node near coord/.append style={font=\footnotesize, color=black},
            ytick=\empty,
            xtick pos=bottom,
        ]
        \addplot coordinates {(1, 12574) (2, 3563) (3, 1925) (4, 1480) (5, 1625) (6, 911) (7, 188) (8, 10) (9, 3)} 
        [fill=frenchblue];
        \end{axis}
    \end{tikzpicture}
    \caption{Distribution of patients by number of hospital visits.}
    \label{dataset:HCTPvisit_dist_hist}
\end{figure}

\subsubsection{Data Annotation}
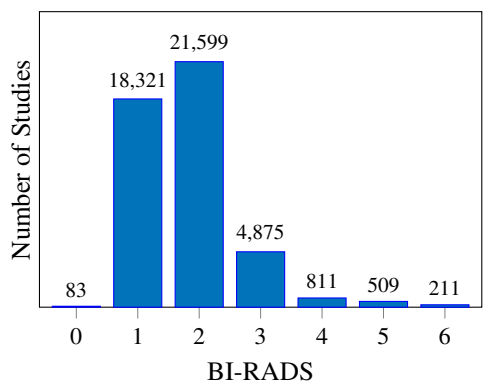
\begin{figure}[t]
    \centering
    \begin{tikzpicture}
        \begin{axis}[
            ybar,
            symbolic x coords={0,1, 2, 3, 4, 5, 6},
            xtick=data,
            ymin=0,
            ymax=26000,
            xlabel={BI-RADS},
            ylabel={Number of Studies},
            ylabel near ticks,
            xticklabel style={font=\small},
            width=0.45\textwidth,
            height=5.5cm,
            bar width=18pt,
            nodes near coords,
            every node near coord/.append style={font=\footnotesize, color=black},
            ytick=\empty,
            xtick pos=bottom,
        ]
        \addplot coordinates {(0, 83) (1, 18321) (2, 21599) (3, 4875) (4, 811) (5, 509) (6, 211)} 
        [fill=frenchblue];
        \end{axis}
    \end{tikzpicture}
    \caption{Distribution of BI-RADS scores per study in the HCTP dataset.}
    \label{dataset:HCTPdist_hist}
\end{figure}

The patient reports are mostly structured and contain the BI-RADS scores assigned to the studies by the radiologists. Based on these reports, BI-RADS scores were extracted for each breast. Additionally, 716 randomly selected studies were manually annotated by four radiologists, and lesion masks were generated for evaluation purposes. One of these radiologists has $35+$ years of experience, two of them have $20+$ years of experience, and one of them was a senior resident in the Department of Radiology at the time of annotation.  BI-RADS distribution of the HCTP dataset is shown in \autoref{dataset:HCTPdist_hist}. These BI-RADS scores were then categorized under three superclasses: negative, benign, and malignant. The negative class includes BI-RADS 1 scores, the benign class includes BI-RADS 2 and BI-RADS 3 scores, and the malignant class includes BI-RADS 4 (4A, 4B, 4C), BI-RADS 5, and BI-RADS 6 scores.

\subsubsection{Training, Validation, and Test Sets}\label{dataset:hctp_stratification}

\begin{figure}[b]
\centering
  \begin{subfigure}{4cm}
    \includegraphics[width=4cm, height=5cm]{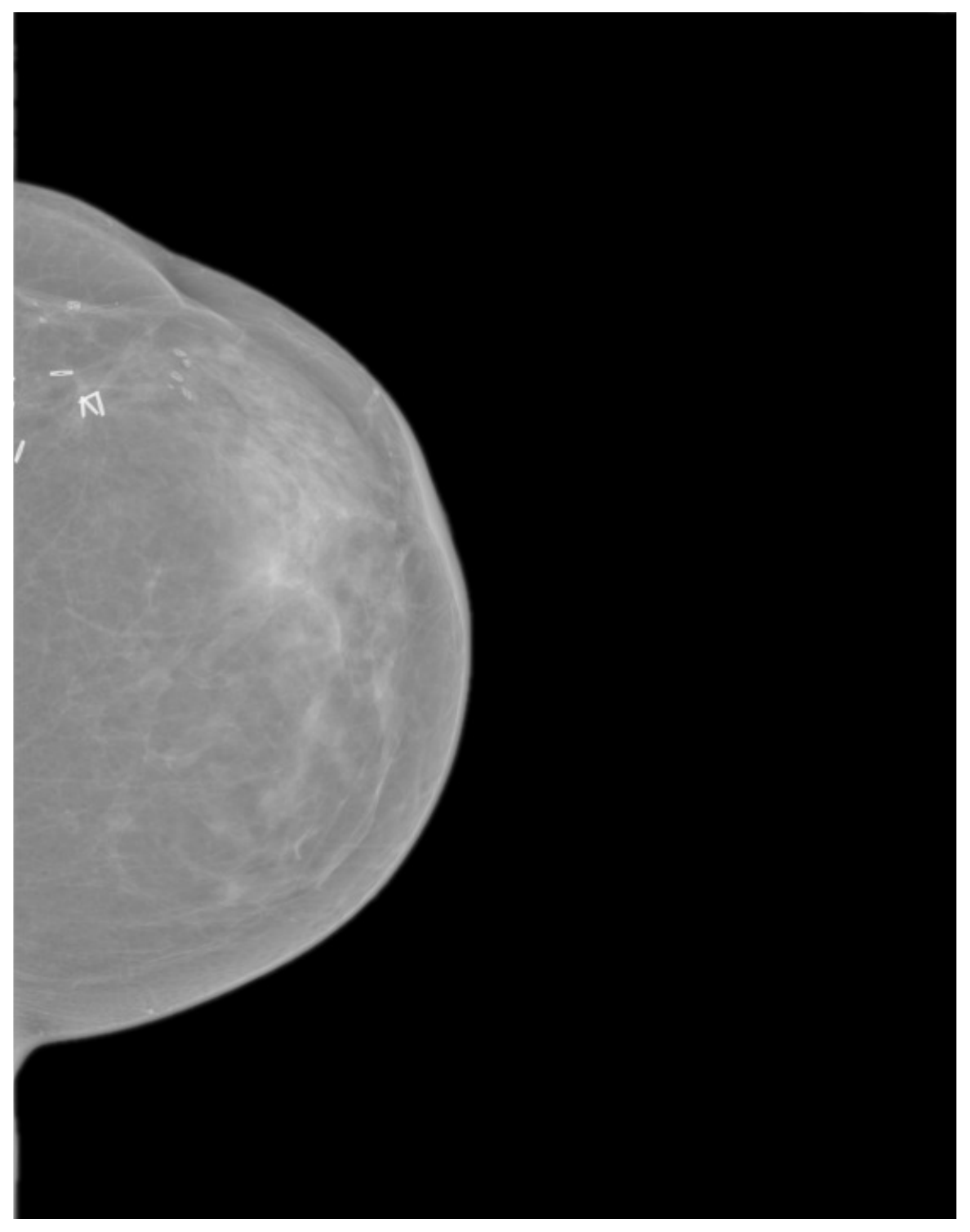}
    \caption{Markers} 
  \end{subfigure}%
  \begin{subfigure}{4cm}
    \includegraphics[width=4cm, height=5cm]{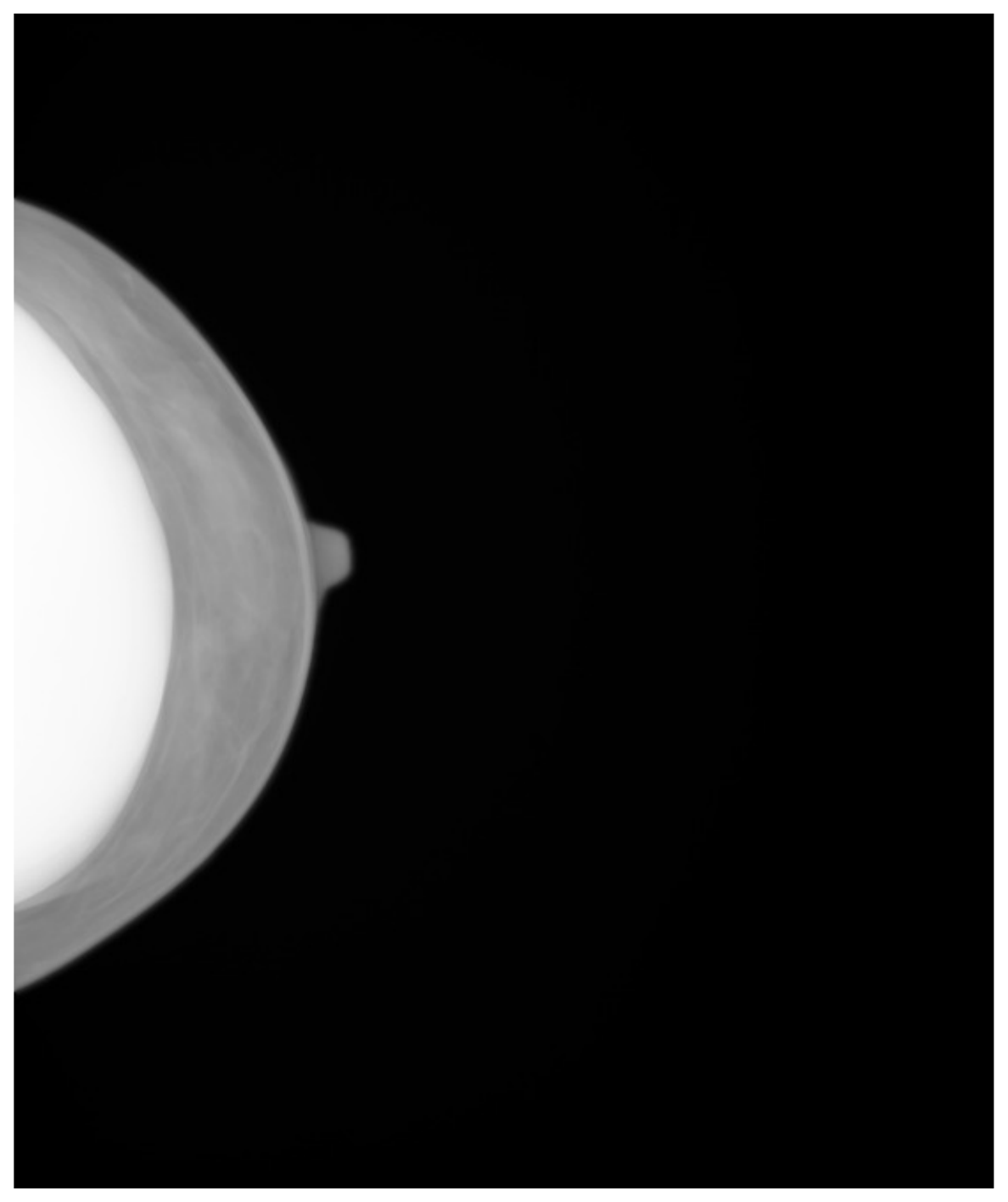}
    \caption{Implant} 
  \end{subfigure}%

\caption{Example images that were excluded from the HCTP dataset.}
\label{fig:HM_RemovedImages}
\end{figure}

The HCTP dataset contains a large corpus of retrospective mammography images, which required substantial effort for filtering and curation. Some images include markers or diagnostic apparatus, while others contain implants, irrelevant objects, or even the patient's or radiologist's hand. Such examples images can be seen in \autoref{fig:HM_RemovedImages}. Our exploratory data analysis (EDA) revealed that part of these cases could be detected using DICOM metadata, whereas others would require custom image-processing techniques, such as histogram-based filtering. To ensure fair model comparison and avoid underestimating model performance, we manually curated a high-quality subset of the HCTP dataset instead of developing such automated filtering algorithms. 

During manual curation, malignant cases with BI-RADS 4 scores lacking pathological confirmation or radiologist annotations were excluded. All patients with malignant findings were then identified, and an equal number of patients with benign findings were randomly sampled. To maintain class balance, an equivalent number of patients diagnosed as negative were randomly selected to match the combined total of malignant and benign cases. Each selected image was manually inspected to confirm that only valid breast regions were retained. If any image within a study contained artifacts or irrelevant objects, the entire CC–MLO pair was discarded. Consequently, each breast was represented by exactly one CC and one MLO view.

The curated subset was stratified to training, validation, and test sets. For stratification, we handled classes separately. Specifically, 80\% of malignant CC–MLO pairs were assigned to the training set, 10\% to the validation set, and 10\% to the test set. The same proportions were applied to benign CC-MLO pairs of similar size. An equal number of negative samples, corresponding to the combined count of benign and malignant samples in the training set, were also included in the training set for diversity.

Some patients presented only unilateral findings (benign or malignant in one breast, negative in the other) or had follow-up studies over multiple years. Such cases can cause data leakage if not carefully managed. To avoid this, all CC–MLO pairs belonging to a given Patient ID were restricted to a single split (training, validation, or test). If any Patient ID was found across multiple splits, all associated images were reassigned to a single split to ensure exclusivity.

Finally, benign and malignant samples were balanced in the test set to standardize the PR-AUC baseline at 0.50 and enable consistent comparison across datasets. After this procedure, the training, validation, and test subsets contained 6,812, 298, and 352 mammography images, respectively, from 1,561, 111, and 139 unique patients.(\autoref{dataset:dist}) Note that a single patient may have multiple studies.

\subsection{VinDr Dataset}
The VinDr is a digital mammography image dataset introduced by \citep{nguyen2023vindr}. The dataset contains 20,000 mammography images from 5,000 studies in monochrome 1 and 2 formats. The images were acquired with the Siemens, IMS, and Planmed mammography machines between 2018 and 2020. Among the monochrome 2 images, 94\% are acquired using Siemens scanners. The dataset also contains BI-RADS scores, finding types like skin thickening and mass, and breast density information. Yet, the authors also indicate that the malignant findings are not confirmed with pathological reports. The authors also provide training and test sets to facilitate comparisons between different research. The provided training set consists of 16,000 mammography images, while the test set includes 4,000 images.

In this study, images in monochrome 1 format were excluded from both the training and test sets. The remaining training set included only 664 malignant samples, and the test set 168 malignant samples, leading to a significant class imbalance. To address this, 1,500 randomly selected negative CC-MLO image pairs and 750 benign image pairs were retained. For the test set, negative samples were excluded, and 250 benign images were retained. After sampling, 15\% of the training samples, selected based on patient IDs, were assigned to the validation set. The number of benign and malignant samples was also balanced for the test set to standardize the PR-AUC baseline scores at 0.50 for the VinDr dataset.

\newcolumntype{C}{>{\centering\arraybackslash}X}

\begin{table*}[t]
    \captionof{table}{Distribution of mammography images across the training, validation, and test sets for HCTP, VinDr, and CSAW datasets.}
    \centering
    \begin{tabularx}{\textwidth}{lCCCC}
        \hline
        \thead{Dataset} &    & \thead{Training} &\thead{Validation} & \thead{Test} \\ 
        \hline
        \multirow{4}{*}{\textbf{HCTP}}      & Negative         & 3760 &  -  &  -    \\ 
                                            & Benign           & 1700 & 128 & 176   \\
                                            & Malign           & 1352 & 170 & 176   \\
                                            & \textbf{Total}   & 6812 & 298 & 352   \\
        \hline
        \multirow{3}{*}{\textbf{VinDr}}     & Negative         & 1500 &  -  &  -    \\ 
                                            & Benign           & 662  & 88  & 168   \\
                                            & Malign           & 564  & 100 & 168   \\
                                            & \textbf{Total}   & 2726 & 188 & 336   \\
        \hline
        \multirow{2}{*}{\textbf{CSAW}}      & Negative/Benign  & 2038 & 256 & 106    \\ 
                                            & Malign           & 838  & 104 & 106    \\
                                            & \textbf{Total}   & 2876 & 360 & 212    \\
        \hline    
    \end{tabularx}
    \label{dataset:dist}
\end{table*}

\subsection{CSAW Dataset}
The CSAW (mammography) dataset contains 24,697 studies from 8723 patients from the Karolinska University Hospital, Stockholm \citep{Strand2022CSAW-CC}. Among these studies, 873 are malignant, while the rest are negative or benign. The collected data covers 98,788 mammography images from 2008 to 2015. All the mammography images were acquired with a HOLOGIC, Inc. mammography machine. 

In this study, screen-detected cancer cases were only included. The dataset does not provide specific labels for samples classified as negative or benign, so these were grouped into a single "negative/benign" category. To address the class imbalance, the negative/benign samples were sampled at twice the number of malignant cases for the training subset. The training set includes a total of 2,876 mammography images, while the validation 360, and the test set contains 212 mammography images \autoref{dataset:dist}.

\subsection{Image Pre-processing and Data Augmentations}
To feed the model, all mammography images were standardized with a set of pre-processing techniques. The images were flipped to the left side, and the background was cropped, with all background pixels set to 0. The images were then resized to 2944×1920 and standardized to have zero mean and unit variance.

During the preprocessing step, it was observed that the histograms of some VinDr images were compressed into a narrow range after normalization. Upon investigation, it was found that these images had histograms covering a larger pixel range, with values shifted to higher pixel intensities.

To address this issue, images with pixel values exceeding 5,000 were adjusted to match the histogram distribution of the remaining images. This was achieved by first detecting the breast region and setting pixel values outside this region to 0. Next, the 1st percentile pixel value of each image was subtracted from all pixel values, and the maximum pixel values were clipped to the 99.9th percentile value of the image.

\begin{figure}[t]
\centering
  \begin{subfigure}{0.16\textwidth}
    \includegraphics[width=\textwidth, height=5cm]{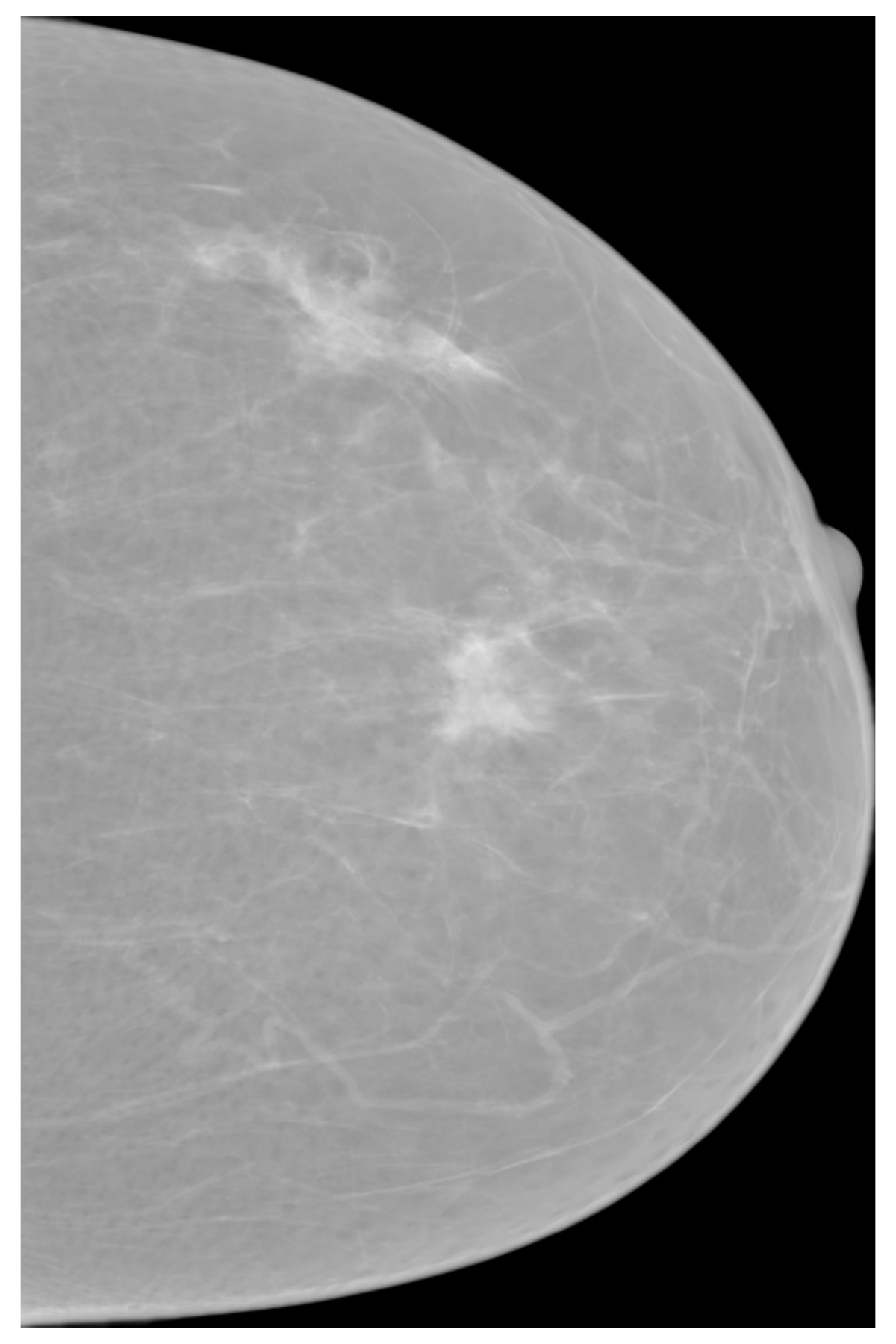}
    \caption{HCTP}
  \end{subfigure}%
  \begin{subfigure}{0.16\textwidth}
    \includegraphics[width=\textwidth, height=5cm]{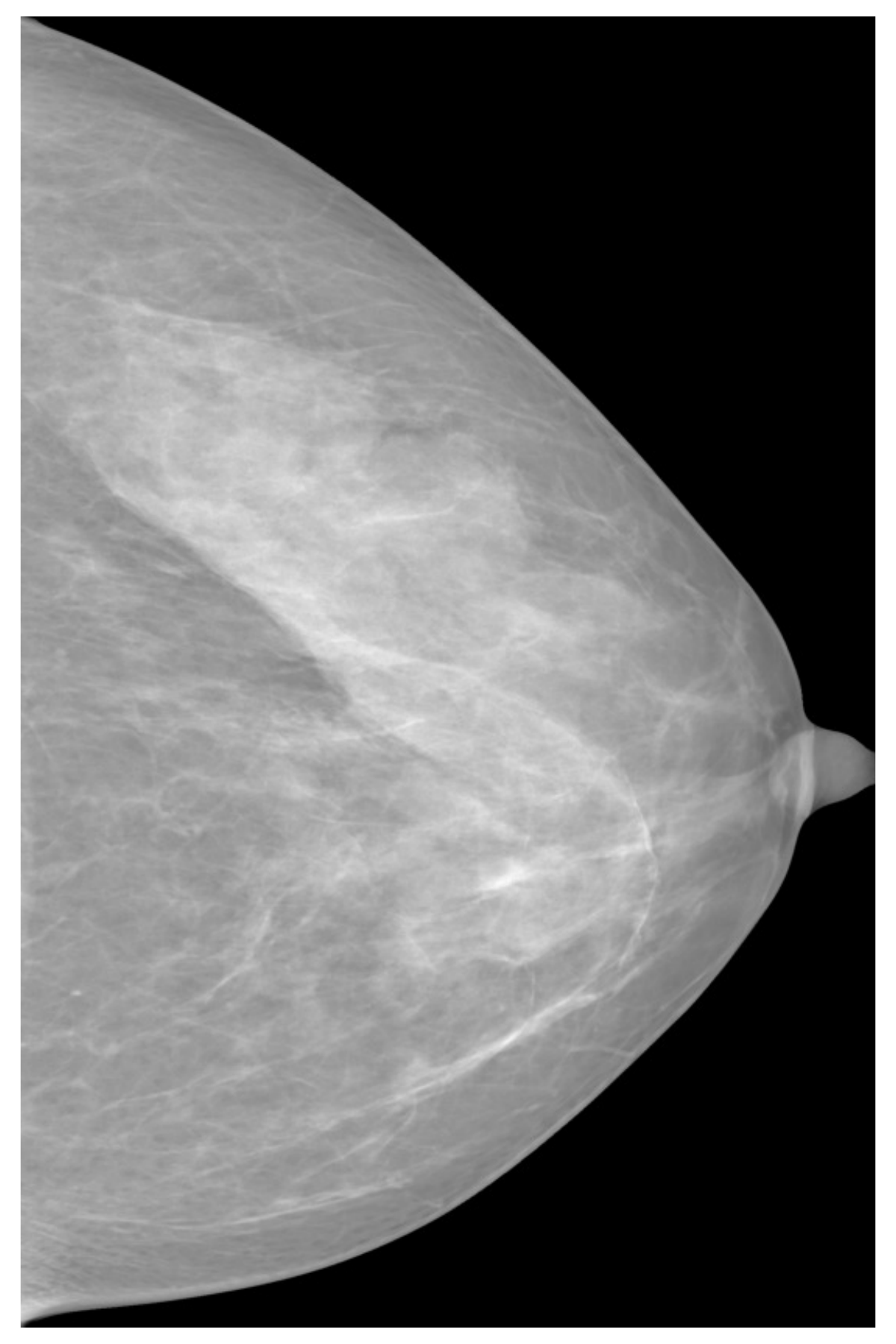}
    \caption{VinDr} 
  \end{subfigure}%
  \begin{subfigure}{0.16\textwidth}
    \includegraphics[width=\textwidth, height=5cm]{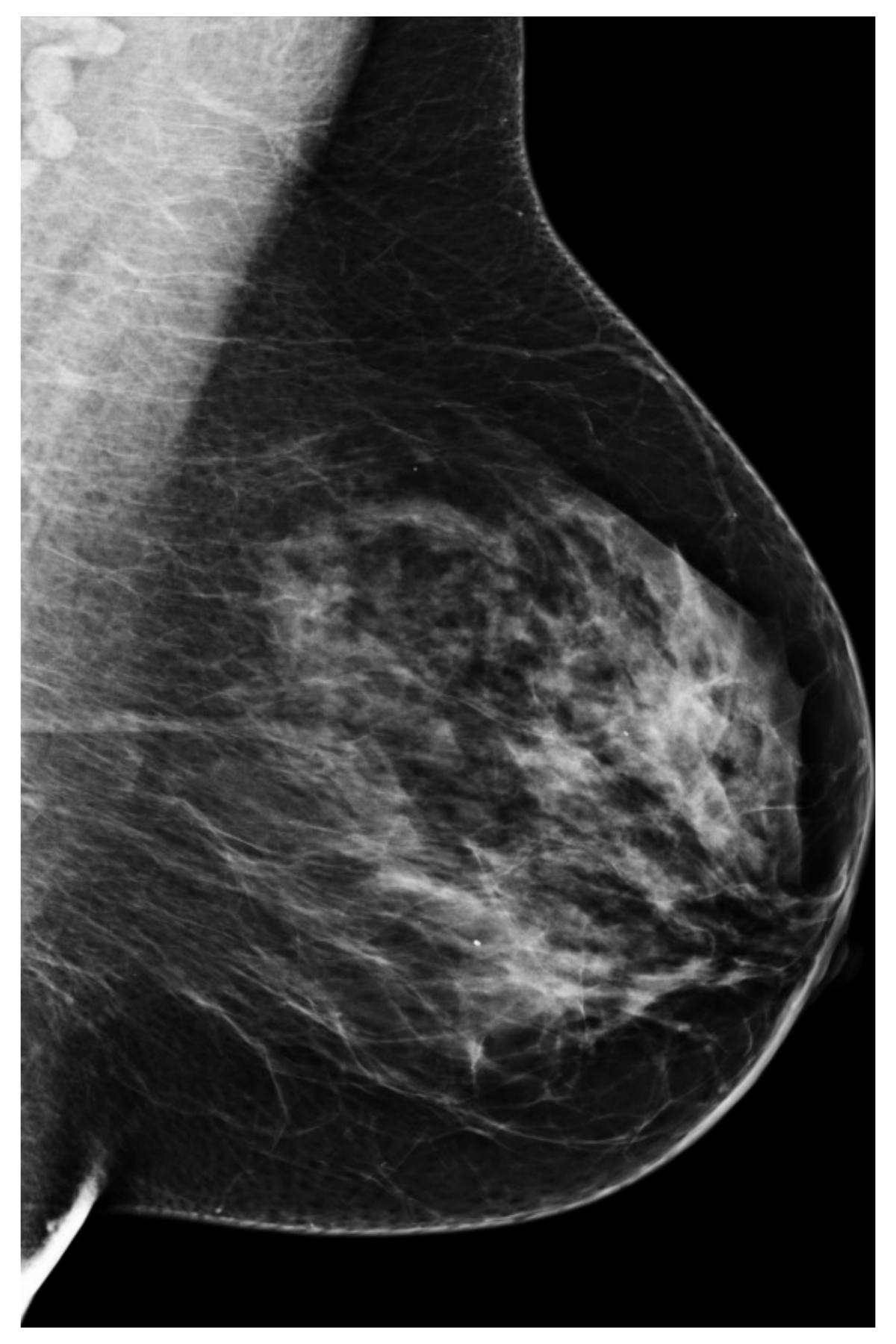}
    \caption{CSAW} 
  \end{subfigure}%
\caption{Standardized mammography samples from Hacettepe, VinDr, and CSAW dataset.}
\label{fig:preprocessedsamples}
\end{figure}

As data augmentation, random Gaussian noise ($\mu=0$, $\sigma=0.005$), random horizontal flipping ($p=0.5$), random rotation ($-15^\circ$, $+15^\circ$), random translation (0, 0.1), random shearing ($-25^\circ$, $+25^\circ$), scaling by a random factor between 0.8 and 1.6 were used. Three standardized mammography images from different datasets are shown in the \autoref{fig:preprocessedsamples}.

\section{Methods}
In this study, we introduce DoSReMC, a framework that addresses the impact of domain shift on mammography classification by specifically adapting BN and FC layers. Unlike prior studies that address domain shift from data and learning perspectives \citep{kumar2017cross, yala2021toward, lauritzen2023robust, quintana2024contrastive}, we analyze its impact from an architectural viewpoint. Specifically, we explore how domain-specific image statistics influence BN layers, leading to performance degradation on unseen domains. 

To further dissect this relationship, we conduct experiments in which only the BN and FC layers are trained, enabling us to isolate the influence of convolutional and normalization layers on model robustness. While similar issues have been studied in other fields \citep{lim2023ttn, li2016revisiting, schneider2020improving, karani2018lifelong, frankle2020training}, to the best of our knowledge, this is the first work to examine the role of BN layers in mammography classification under domain shift. 

Additionally, we evaluate the effectiveness of DoSReMC combined with partial DAT in mitigating domain-shift effects arising from variations in mammography devices.

\subsection{Batch Normalization (BN)}
BN is a crucial component of modern computer vision architectures, particularly CNNs. \citep{ioffe2015batch} introduced BN to mitigate internal covariate shift (ICS), which refers to the change in the distribution of layer inputs during training. By addressing ICS, BN was initially thought to facilitate faster and more stable convergence. However, \citep{santurkar2018does} argued that the effectiveness of BN does not primarily stem from reducing ICS. In fact, they showed that ICS may even increase with BN. Instead, BN improves optimization by smoothing the loss landscape, which leads to more predictable gradients and accelerates convergence. Complementing this view, \citep{bjorck2018understanding} showed that BN allows for the use of higher learning rates without causing training divergence, which further contributes to faster convergence and improved generalization. While these studies focus on the optimization benefits of BN, in this work, we investigate its role from the perspective of domain shift.

\begin{equation}
    \hat{x}^{(k)} = \frac{x^{(k)} - \mu_{\mathcal{B}}^{(k)}}{\sqrt{\sigma_{\mathcal{B}}^{(k)2}}}, \quad
    y^{(k)} = \gamma^{(k)} \hat{x}^{(k)} + \beta^{(k)}
\end{equation}

Here, $\mu_{\mathcal{B}}^{(k)}$ and $\sigma_{\mathcal{B}}^{(k)2}$ denote the mean and variance of the $k$-th feature over the mini-batch $\mathcal{B}$, respectively. The parameters $\gamma^{(k)}$ and $\beta^{(k)}$ are learnable and allow the network to recover the original distribution if necessary.

In practice, during training, moving averages of the mini-batch mean $\mu_{\mathcal{B}}^{(k)}$ and variance $\sigma_{\mathcal{B}}^{(k)2}$ are computed and then fixed for use during inference. In this work, we are particularly interested in evaluating how these moving averages, obtained from a specific training dataset, perform when applied to other datasets, especially when the training and target datasets exhibit different pixel intensity distributions. While BN enables faster convergence and improved generalizability within the same or similar domains, it raises a critical question: does it increase domain dependency by favoring the training distribution, ultimately degrading performance under domain shift?

\subsection{Model}
To investigate this, we adopt the CNN-based model proposed in \citep{shen2021interpretable} within the DoSReMC framework for targeted BN adaptation. This model was originally trained on a large mammography dataset, the NYU dataset \citep{wu2019nyu}, containing images from Siemens and Hologic scanners, devices that align with those used in the VinDr and CSAW datasets. Given its demonstrated effectiveness in breast cancer recognition, we hypothesize that it would serve as a strong baseline. However, our analysis is not confined to this specific architecture; the insights we present are applicable to other models that incorporate BN layers.

\begin{figure*}[t]
\centering
    \includegraphics[width=\textwidth]{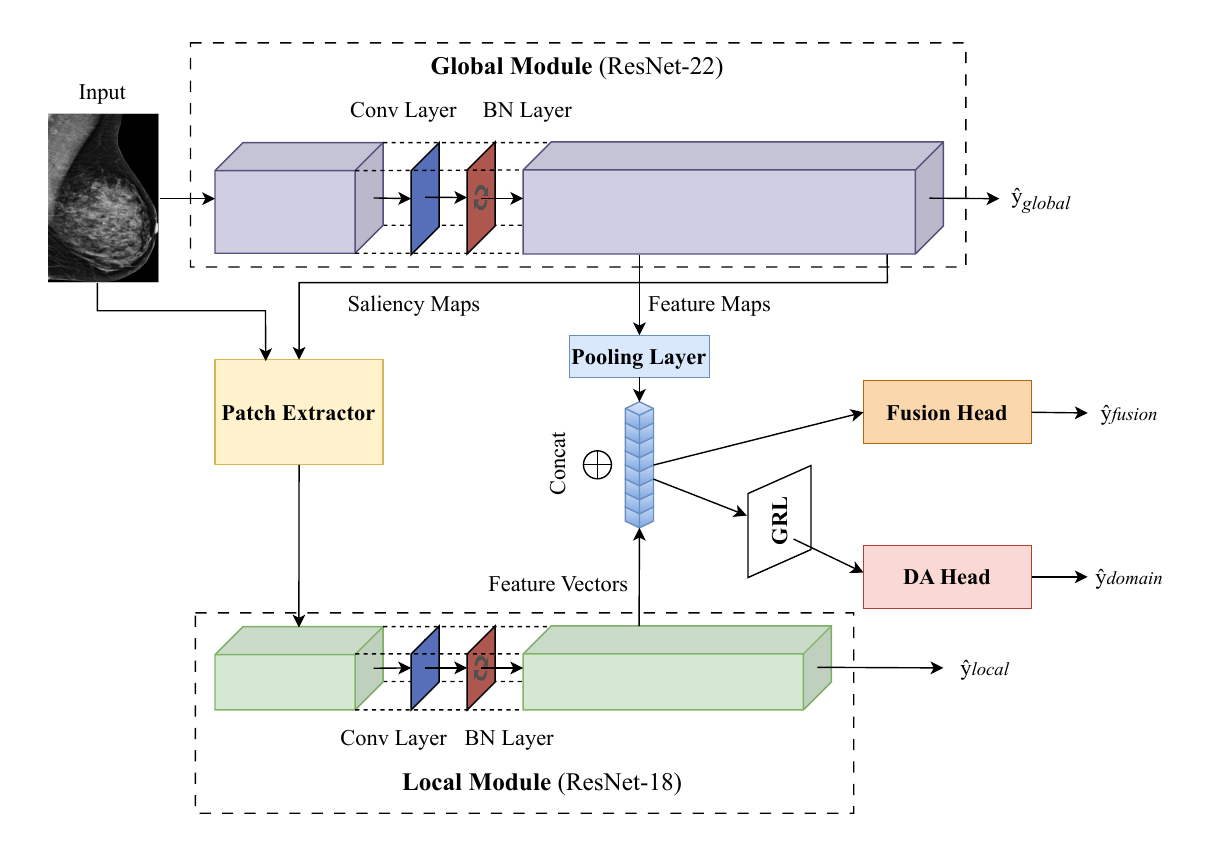}
\caption{The figure illustrates the two-step architecture proposed in \citep{shen2021interpretable}, which combines global and local information, adapted here within a DAT setting. The global module utilizes a ResNet-22 model, which includes one additional residual block and a quarter of the filters in each convolutional layer compared to the ResNet architectures \citep{he2016deep}, while the local module employs a ResNet-18 model. The GRL, implemented as described in \citep{ganin2016adversarial}, multiplies the gradient by a negative constant ($\lambda_{\text{domain}}$) during backpropagation while acting as an identity transformation during forward propagation. The final classification of an image as benign or malignant is determined by the fusion module. The adversarial module classifies whether the image belongs to the source or target domain. The Domain Adversarial module and GRL are used only for the domain adversarial experiments.}
\label{model:architecture}
\end{figure*}

The proposed model comprises three sub-modules, and we modify it for domain adversarial training, as shown in \autoref{model:architecture}. The global module of the proposed model processes high-resolution mammography images and identifies suspicious regions using saliency maps, and to reduce computational cost, it utilizes a low-capacity ResNet-like architecture. The patch extractor then generates patches from the saliency maps based on a predefined size. These patches contain more detailed information about the suspected regions, aiding in the differentiation between benign and malignant findings. To enhance this analysis, a high-capacity local network processes the patches, extracting feature vectors that represent the regions of interest.

Finally, the fusion module integrates the feature maps from the global module with the attention-weighted feature vectors from the local module to produce the final prediction in a multi-label format.

\subsection{Domain-Adversarial Training}
While the main mechanism of DoSReMC is selective BN and FC fine-tuning, we also incorporate partial DAT \citep{ganin2016adversarial} to suppress domain-specific signals in BN layers and align contextual features across domains. In this setting, we apply DAT only to BN and FC layers, freezing convolutional layers, thereby preserving pretrained representations while encouraging BN statistics that generalize across domains.

In our DAT setup, the domain head acts as an adversary to the global, local, and fusion modules of the model. While the domain head attempts to classify the domain of the input image, the global and local modules are trained to confuse it. Simultaneously, the global, local, and fusion prediction heads perform benign versus malignant classification, guiding the feature extraction modules to focus on task-relevant, domain-invariant information. The underlying assumption is that, if the domain head fails to distinguish domains, the model has successfully discarded domain-specific representations in favor of contextual features.

To achieve this, we implement a Gradient Reversal Layer (GRL) and a domain head, as depicted in \autoref{model:architecture}. During forward propagation, the GRL acts as an identity function, leaving activations unchanged. However, during backpropagation, it reverses the gradients, disrupting domain-specific features that help distinguish domains and thereby confusing the domain head. The domain head consists of three fully connected layers with 768, 256, and 128 neurons, respectively, each followed by a ReLU activation. The output layer contains two neurons with a softmax activation, predicting whether an image belongs to the source or target domain. The complete implementation of the model can be found in the repository\footnote{\label{model:repo}\githubrepo}. 

Overall, this training strategy forces the model to remove domain-specific features, encouraging the domain head to misclassify domains while enabling the fusion head to focus on distinguishing between benign and malignant findings.

\subsection{Loss function}
The classification loss function described in \citep{shen2021interpretable} is employed to train the model, as shown in \autoref{eq:gmic_loss}. The total classification loss is computed by summing the outputs of the global, local, and fusion modules, along with the L1 norm of the saliency maps (SM), enabling end-to-end training. This function is defined as follows:

\begin{equation}
    \begin{aligned}
        \ell_{c}(y, \hat{y}) = \sum_{c \in \{b,m\}} &\ \text{BCE}(y^c, \hat{y}^c_{\text{local}}) 
         + \text{BCE}(y^c, \hat{y}^c_{\text{global}}) \\
        & + \text{BCE}(y^c, \hat{y}^c_{\text{fusion}}) 
        + \beta L_{\text{reg}}(\mathbf{SM}^c) 
    \end{aligned}
    \label{eq:gmic_loss}
\end{equation}

where BCE denotes  Binary Cross Entropy, and $\beta$ controls the regularization strength of the saliency maps. The variable $c \in \{b, m\}$ indexes the classes in the multi-label setting, corresponding to \textit{benign} and \textit{malignant}, respectively. Here, $y^c$ represents the ground truth labels, while $\hat{y^c}$ denotes the predictions from the $local$, $global$, and $fusion$ heads.

To train the model in a domain adversarial setting, the domain loss term is defined as follows:
\begin{equation}
    \begin{aligned}
        \ell_{d}(y, \hat{y}) =  \sum_{d \in \{s,t\}} &\ \text{BCE}(y^d, \hat{y}^d_{\text{domain}})
    \end{aligned}
    \label{eq:da_loss}
\end{equation}

where $y^d$ represents the ground truth domain label (source or target), and $\hat{y^d}$ denotes the corresponding prediction from the domain head.

The total loss used to train the model end-to-end is calculated as follows:
\begin{equation}
    \begin{aligned}
        \mathcal{L}(\ell_{c}, \ell_{d}) =  \ell_{c} + \ell_{d}
    \end{aligned}
    \label{eq:total_loss}
\end{equation}

\newcommand{\expsetup}[3]{\mathcal{M}_{\mathrm{#1} \rightarrow \mathrm{#2}}^{\mathrm{#3}}}
\newcommand{\model}[2]{\mathcal{M}_{\mathrm{#1}}^{\mathrm{#2}}}

\section{Experiments}
\label{section:experiments}

\begin{figure}[b]
\centering
    \includegraphics[width=7.5cm, height=5.5cm]{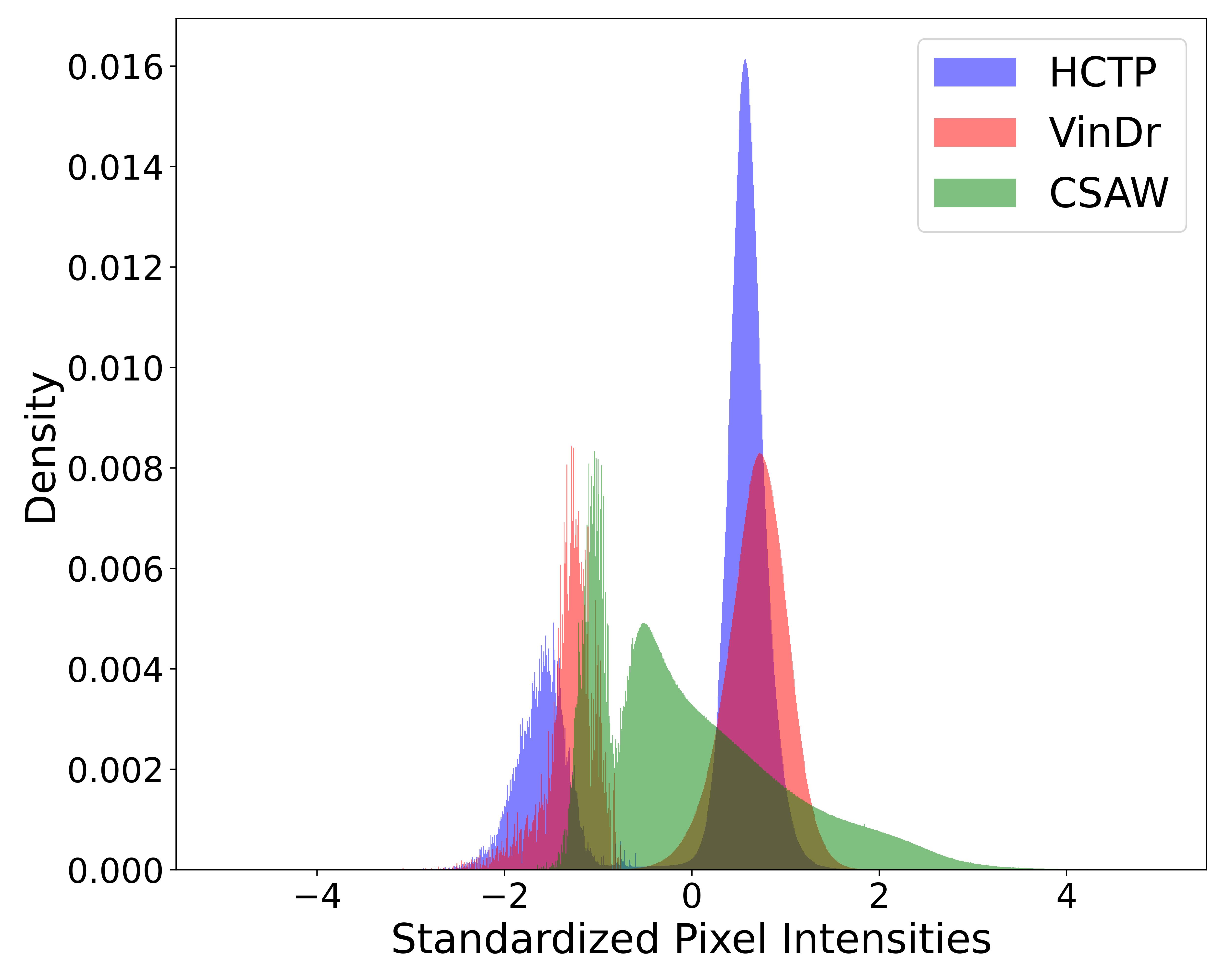}
\caption{Pixel-intensity distributions of mammography images from HCTP, VinDr, and CSAW datasets.}
\label{fig:dataset_hists}
\end{figure}

Our hypothesis is based on the impact of statistical variations in image intensity distributions and how these affect model generalization across datasets through the behavior of BN layers. In our study, we observed noticeable shifts in pixel intensity distributions across the HCTP, VinDr, and CSAW datasets, as shown in \autoref{fig:dataset_hists}. Although full access to the acquisition protocols is unavailable, the observed discrepancies are likely due to differences in imaging devices or device-specific image processing techniques, a hypothesis supported by prior research \citep{jimenez2023memory}.

To better understand the impact of these distributional shifts, we further investigate how BN layers respond to domain-specific variations and how they influence overall model performance. To this end, we fine-tune models and evaluate them with and without using the moving averages computed during training. In the latter case, we employ test-time BN, recomputing batch statistics during inference using batch sizes of 8 and 64. The substantial increase in batch size, 8 to 64, is intentional and aimed at investigating how batch size influences test-time performance. Additionally, we explore DAT as a potential remedy for the performance degradation caused by domain-specific shifts. Unlike the conventional approach of tuning all model layers with DAT, we exclusively apply it only to the BN and FC layers, which distinguishes our approach from the conventional ones.

We designed a series of experiments to systematically investigate the performance and behavior of our model under various conditions. 

Throughout this section, we use the notation $\model{source}{statistics}$ to denote models, where \textit{source} refers to the training dataset and the superscript \textit{statistics} indicates the BN statistics applied during inference. The extended notation $\expsetup{source}{target}{statistics}$ represents an evaluation setup in which a model trained on the \textit{source} dataset is assessed on the \textit{target} dataset. Specifically, superscript \textit{tr} refers to moving averages of BN statistics computed during training (i.e., training-time statistics), while \textit{tt} indicates statistics recomputed on the test batch at inference time (i.e., test-time statistics).

\begin{itemize} 
    \item First, we evaluate the performance of the model trained exclusively on the NYU dataset ($\model{NYU}{}$), on the HCTP, VinDr, and CSAW datasets, respectively, without any fine-tuning. Since the NYU dataset includes images acquired from both Siemens and Hologic devices, it enables us to assess the model's generalization across similar domains. The CSAW dataset contains images exclusively from Hologic devices, while VinDr primarily includes images from Siemens devices. In contrast, the HCTP dataset, which consists of images from GE devices, provides insight into the model’s behavior on out-of-domain data.

    \item Second, we intentionally alter the input distribution by normalizing input images to the range [0, 1], and evaluate model performance using both training- and test-time BN statistics. The goal of this controlled experiment is to observe how an artificially induced distribution shift affects model performance. This experiment, denoted with an apostrophe (’), is represented as $\model{NYU}{'tr}$ and $\model{NYU}{'tt}$.
    
    \item Third, we extend our investigation to real-world distribution shifts by fine-tuning the NYU pretrained model separately on the HCTP dataset, the VinDr dataset, and their combined version (HCTP+VinDr). We then evaluate each of these fine-tuned models on the respective unseen dataset (a cross-dataset evaluation) to assess the impact of dataset shifts on the performance. This experiment also serves to estimate the upper-bound performance achievable through transfer learning on the source domains. These models are denoted as $\model{HCTP}{}$, $\model{VinDr}{}$, and $\model{HCTP+VinDr}{}$, respectively. We excluded the CSAW dataset from fine-tuning due to potential label noise stemming from publicly available labels limited to malignant and negative cases, with benign labels unavailable. However, we retained CSAW as an evaluation dataset to assess model performance on unseen data.

    \item Fourth, we incorporate the DoSReMC approach by freezing the convolutional layers and fine-tuning only the BN and FC layers using the HCTP, VinDr, and HCTP+VinDr datasets. These models, denoted as $\model{HCTP(BNFC)}{tr}$, $\model{VinDr(BNFC)}{tr}$, and $\model{HCTP+VinDr(BNFC)}{tr}$, aim to evaluate whether adapting the learnable parameters of BN layers, namely the scaling ($\gamma$) and shifting ($\beta$) parameters, along with the first- and second-order statistics—could lead to performance improvements on the target domains.

    \item Finally, we extend DoSReMC with DAT by training a model on the HCTP+VinDr dataset to improve the model's generalizability. We aim to remove domain-specific statistics accumulated by BN layers in order to achieve consistent performance across HCTP, VinDr, and CSAW datasets. In this setup, we apply DAT only to the BN and FC layers, freezing the convolutional layers ($\model{\text{HCTP+VinDr(DA\_BNFC)}}{}$). The HCTP+VinDr dataset serves as the source domain, while CSAW was used as the target domain. Only domain labels (source vs. target) from the CSAW dataset are used to train the domain discriminator; no malignant or benign class labels from CSAW were included in the training process. For comparison, we also report results obtained by applying DAT to all layers, represented by $\model{\text{HCTP+VinDr(DA)}}{}$, including the convolutional layers.

\end{itemize}

\subsection{Hyper-parameters}
For consistency with the research associated with the proposed method, the same set of hyperparameters described by \citep{shen2021interpretable} is applied across all experiments. We parameterize the global module using the ResNet-22 architecture as described in \citep{shen2021interpretable}, while we parameterize the local module using ResNet\-18. We conduct the experiments using the Adam optimizer with a momentum rate of 0.001. The model weights are initialized with a pre-trained model (\textit{sample\_model\_1.p}) available in the GitHub repository \footnote{\url{https://github.com/nyukat/GMIC/tree/master/models} (Retrieved on 14 August 2025) \label{link:gmic}} associated with the research. 

We optimize the learning rate ($\eta$) and regularization weight ($\beta$) via a log-uniform random search with Ray Tune \citep{liaw2018tune}, sampling $\eta$ from $10^{-5.5}$–$10^{-4}$ and $\beta$ from $10^{-5.5}$–$10^{-3.5}$. We set top \textit{t}\% to 0.02, the patch size to $w=h=256$, and the number of patches (\textit{K}) to 6. During training, we compose each batch using benign, malignant, and negative samples in a 1:1:2 ratio to ensure consistent class representation; the batch size is set to 8. 

We train the models for 50 epochs, except for the domain-adversarial training, which is trained for 150 epochs. We do not use learning rate scheduler. The model that achieved the highest Precision-Recall Area Under Curve (PR-AUC) score on the malignant class is selected to evaluate the classification performance.

We conducted all experiments on two NVIDIA GeForce RTX 3090, 24 GB GPUs. Additionally, for evaluations requiring a batch size of 64, NVIDIA A100 GPUs were used. 

\subsubsection{Hyper-parameters for domain adversarial training}
Training a model in a domain adversarial setting is a challenging task that requires careful tuning of hyperparameters. If domain adversarial updates are not properly managed, the model may fail to learn effectively or converge to suboptimal results. To address this, we first form batches by selecting half of the samples from the source domain and half from the target domain. The class distribution in the source domain is determined according to the strategy described in \autoref{dataset:hctp_stratification}. The batch size is set to 16. To mitigate noise and limit the influence of the domain adversarial head in the early stages of training, its contribution can be controlled using $\lambda_{p}$. By initializing $\lambda_{p}$ with lower values and gradually increasing it over time, the model can achieve more stable learning.

In this study, we redefine the scheduler proposed in \citep{ganin2016adversarial} for $\lambda_{p}$ to allow adjustment of its upper limit to any value between 0 and 1. The modified lambda scheduler is defined as follows:
\begin{equation}
    \lambda_{\text{domain}} = \frac{2 \times \tau_{\max}}{1 + \exp(-\gamma .p)} - \tau_{\max}
\end{equation}

where $\tau_{\max}$ represents the upper limit of the scheduler. We empirically search various values between 0 and 1 for $\tau_{\max}$ and ultimately set it to 1 based on the observed outcomes. The variable $p$ denotes the training progress, ranging from 0 to 1, and $\gamma$ is set to 10. Note that in our experiments, the domain loss is neither limited by an upper bound nor assigned any weighting factor.

\subsection{Evaluation Metrics}
In order to evaluate the model's performance, we use the Receiver Operator Characteristic - Area Under Curve (ROC-AUC) and PR-AUC. While evaluating the model, we average the predictions corresponding to CC and MLO images of a breast in the same study, and obtain final prediction values. Each model training is repeated four times with different random seeds, and we report the mean and standard deviation across these runs. For the pretrained NYU models, we average the predictions from the five available pretrained models (see footnote~\ref{link:gmic}). Our primary goal is to achieve low false-positive and false-negative rates for malignant cases.

In addition to ROC-AUC and PR-AUC, we report sensitivity, specificity, and F1-score as threshold-based metrics. For their computation, we first select the model achieving the highest PR-AUC among runs with different random seeds, then determine, for each class separately, the decision threshold that maximizes the F1-score for that model.

To assess model calibration, we include reliability diagrams. These analyses were performed using the same models selected for threshold-based evaluations to ensure consistency.

We also analyze how the output distributions of BN layers vary across different models. For this purpose, we extract the BN layer outputs from each model using the same batch of input samples. We then compute the Jensen–Shannon (JS) divergence for each BN layer to quantify the extent of distributional shifts. Additionally, we visualize the kernel density estimations (KDEs) of these outputs to understand the impact of domain shift better. For analyses of saliency maps, JS divergence computations, and KDEs, we used the first trained model from each training setup. Since all models were trained under identical configurations and showed consistent performance across runs, this model was considered representative for visualization purposes.

\section{Results}
In this section, results marked with $*$ and $**$ indicate evaluations performed using the test-time BN statistics. Specifically, * denotes test-time BN applied with a batch size of 8, and ** denotes test-time BN with a batch size of 64.

\subsection{Model performance without fine-tuning}
The results in the $\model{NYU}{tr}$ row of \autoref{results:table:standardization_malign} show that the model generalized poorly when classifying mammography images from the HCTP and VinDr datasets. In contrast, its performance on the CSAW dataset is remarkably high, achieving a PR-AUC of 0.93. The NYU dataset \citep{wu2019nyu} contains approximately twice as many images acquired from Hologic devices compared to Siemens devices and does not contain images from a GE device. We hypothesize that the model exhibits a vendor-specific bias, resulting in significantly improved performance on images acquired from Hologic mammography systems compared to those from other manufacturers.

\begin{figure}[!t]
  \begin{subfigure}{0.16\textwidth}
    \includegraphics[width=\textwidth, height=5cm]{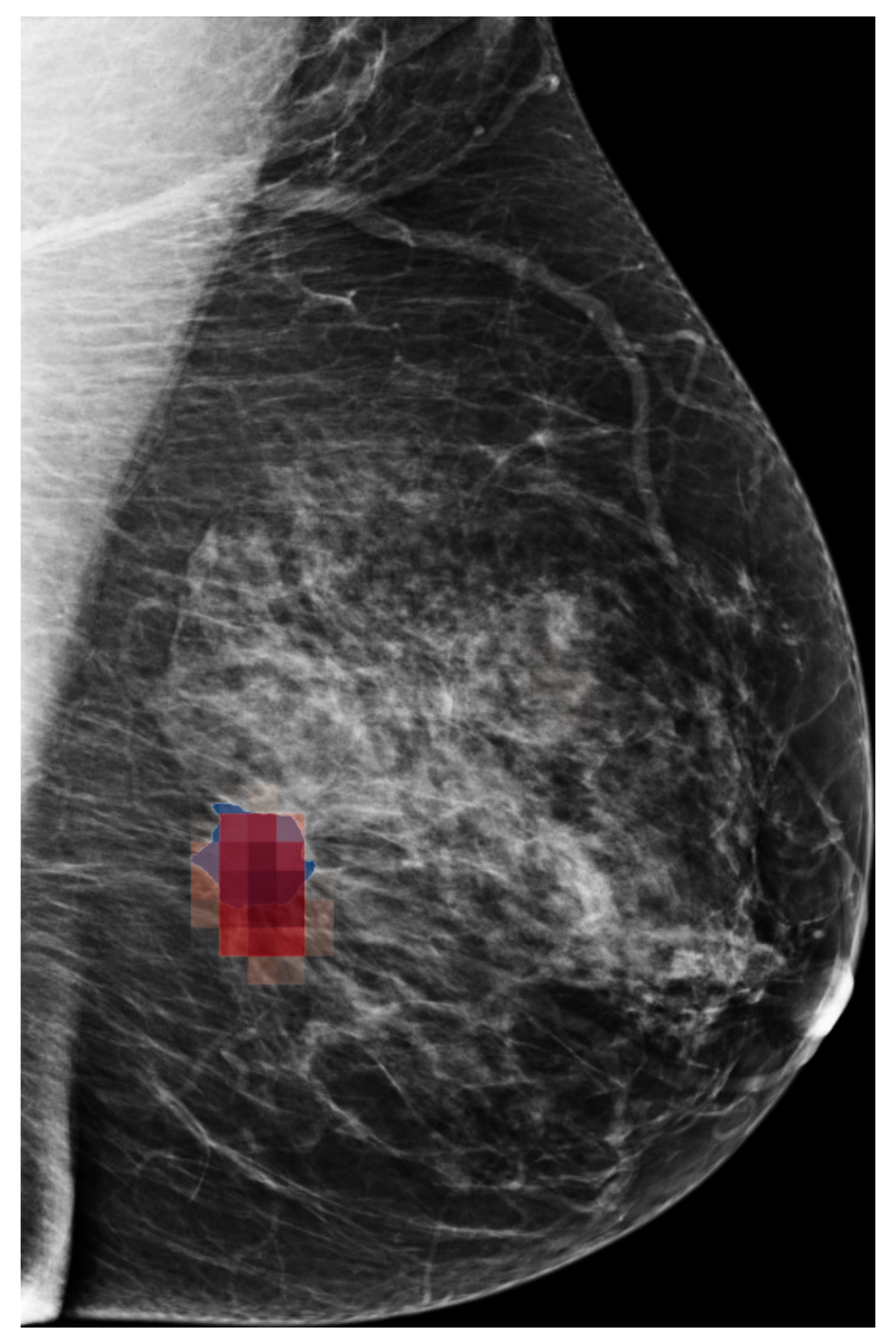}
    \caption{$\model{NYU}{tr}$}
  \end{subfigure}%
  \begin{subfigure}{0.16\textwidth}
    \includegraphics[width=\textwidth, height=5cm]{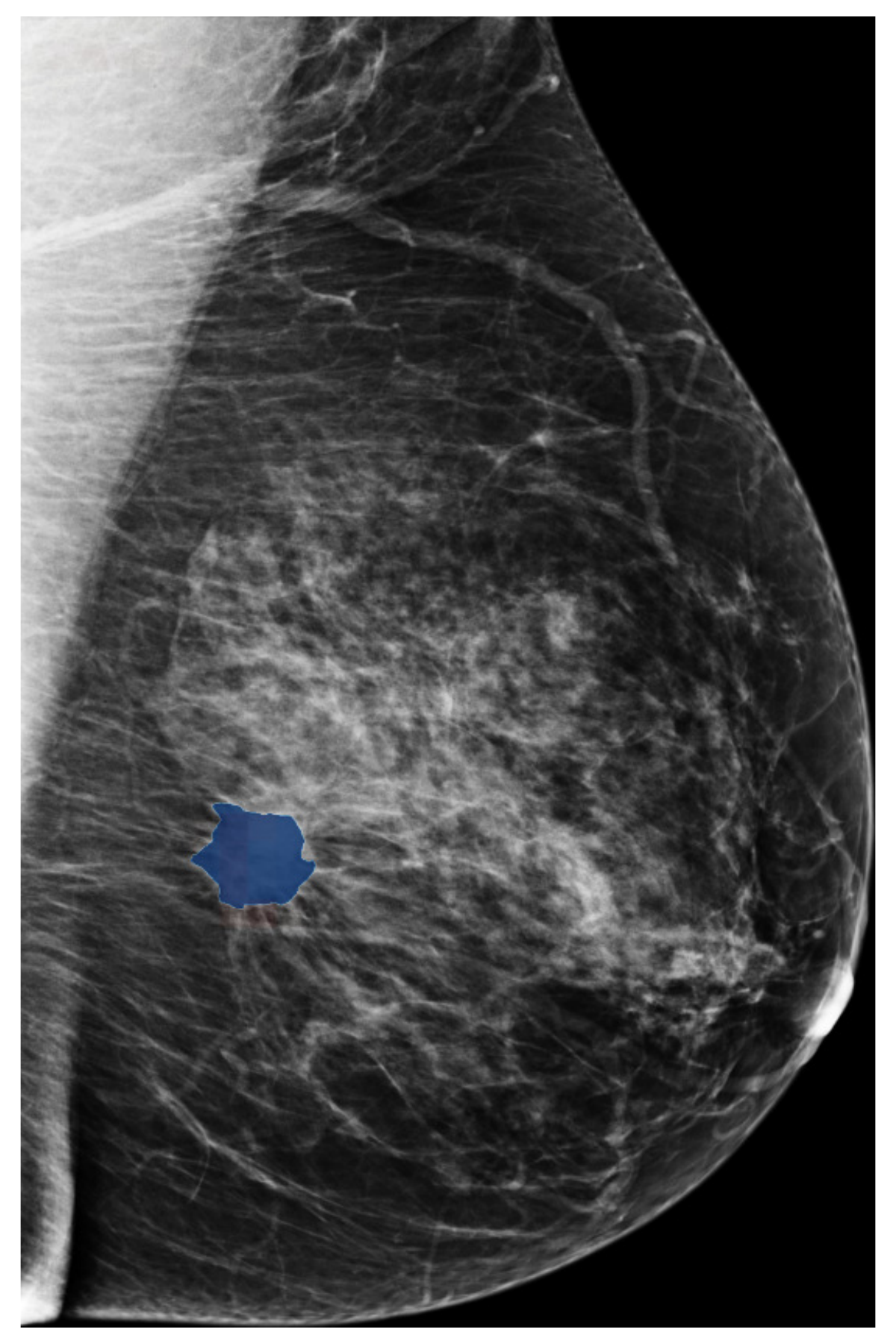}
    \caption{$\model{NYU}{'tr}$} 
  \end{subfigure}%
  \begin{subfigure}{0.16\textwidth}
    \includegraphics[width=\textwidth, height=5cm]{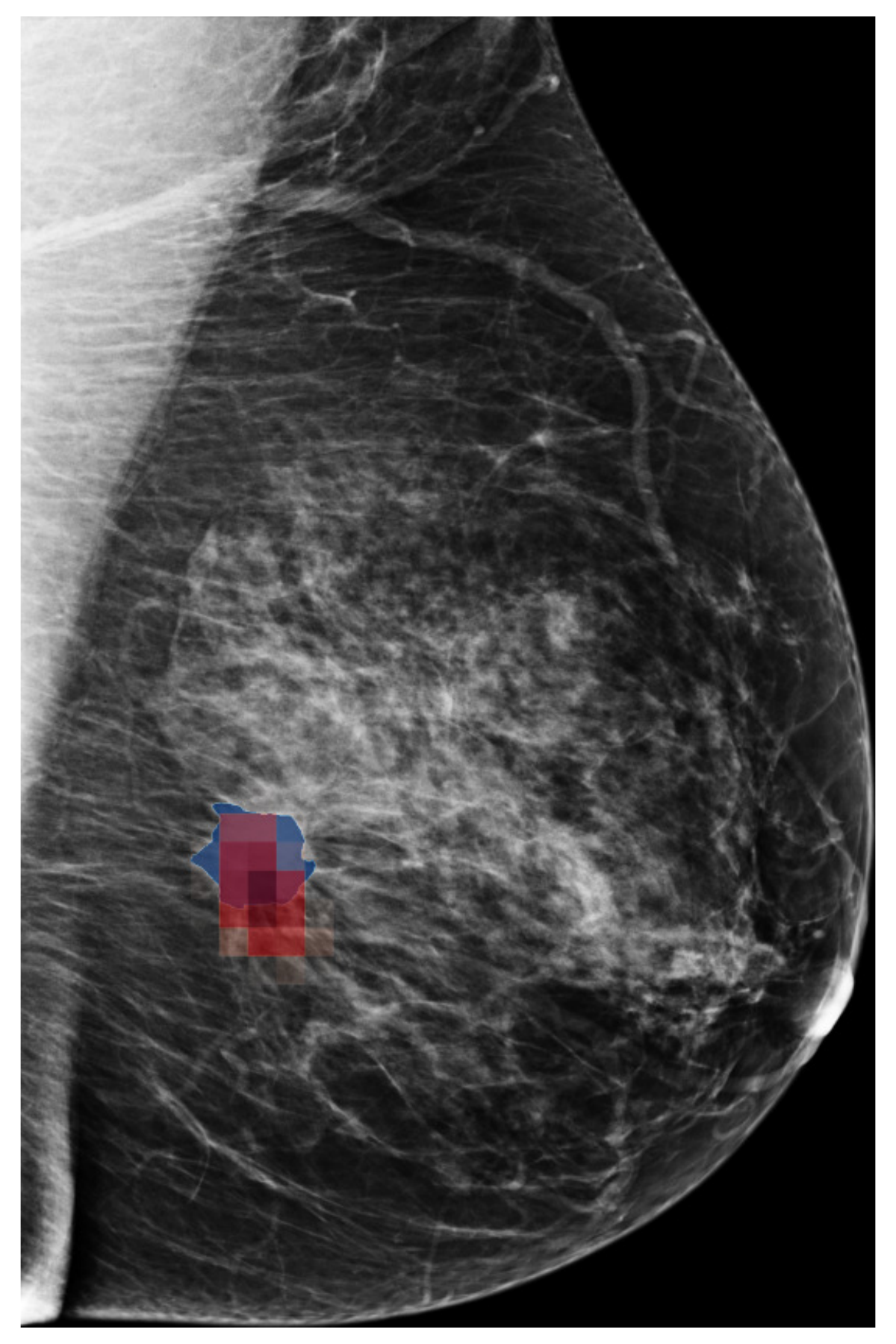}
    \caption{$\model{NYU}{'tt}$} 
  \end{subfigure}%
\caption{Saliency maps generated by $\model{NYU}{tr}$, $\model{NYU}{'tr}$, $\model{NYU}{'tt}$  models on a sample from the CSAW dataset. Blue regions represent the ground truth annotations, while red regions highlight the saliency maps corresponding to the malignant class.}
\label{results:nyu_tr_tt_smaps}
\end{figure}

However, incorporating test-time BN statistics led to a performance improvement on the VinDr dataset, approximately a 13\% improvement in PR-AUC, and yielded only a marginal gain on the HCTP dataset, approximately a 3\% improvement in PR-AUC, while it resulted in a decline in both ROC-AUC and PR-AUC on the CSAW dataset (see the $\model{NYU}{tt}*$ and $\model{NYU}{tt} **$ rows in \autoref{results:table:standardization_malign}). As for the CSAW dataset, the performance drop was expected, given that the model was presumed to have already collected representative population estimates for this domain, making training-time BN statistics more suitable.

\begin{figure*}[!t]
\centering
    \includegraphics[width=14cm, height=6cm]{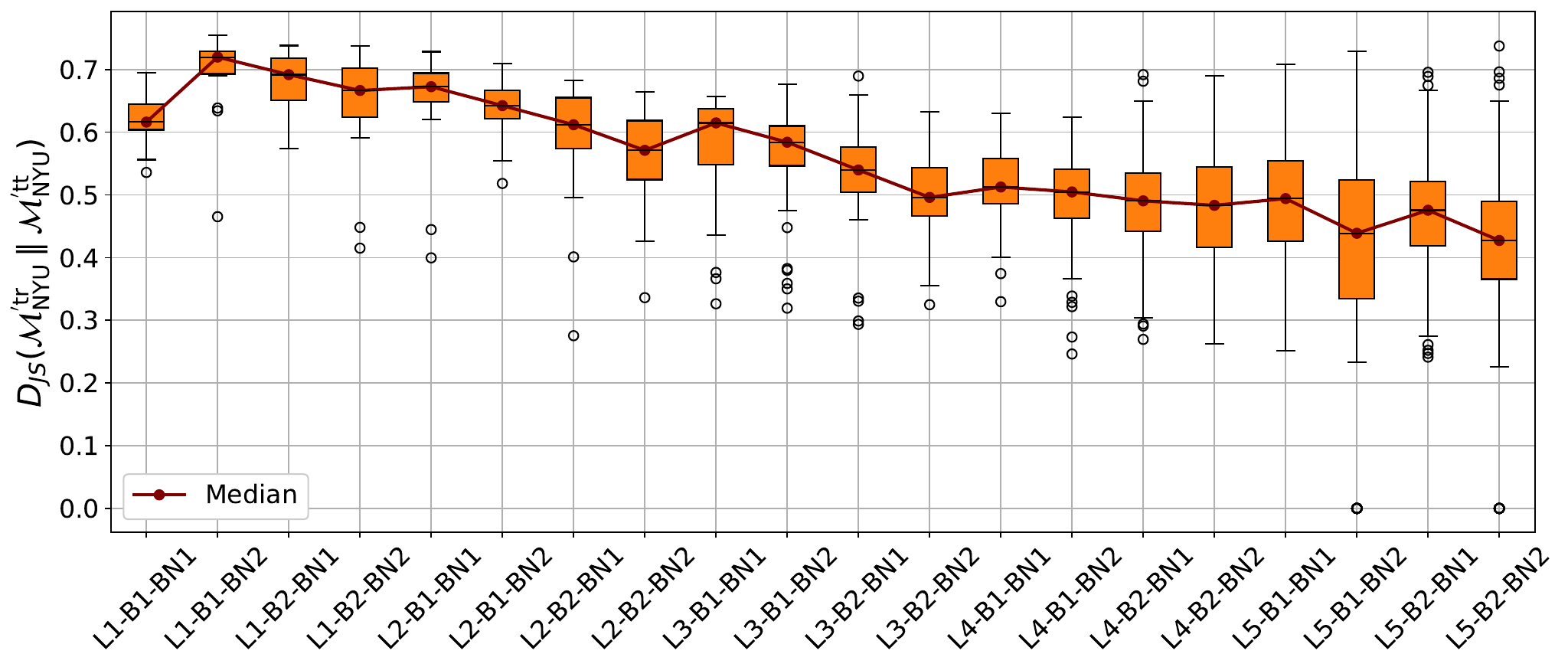}
\caption{JS divergence across BN layers in the Global Module (ResNet-22), using $\model{NYU}{'tr}$ versus $\model{NYU}{'tt}$ on a sample batch from CSAW.}
\label{results:nyu_tr_tt_jsdiv}
\end{figure*}

The VinDr test set used in our experiments consists of approximately 80\% images acquired from Siemens devices. We hypothesize that the observed improvement is due to the fact that the NYU dataset already contains images from Siemens devices. Although the pretrained model’s BN statistics are biased toward Hologic devices, the convolutional layers retain general representational capacity. By updating the BN statistics at test time, the model can better adapt to Siemens images, thereby improving performance on the VinDr dataset under domain shift.

In our second experiment, we investigated the effect of an artificially induced domain shift on the $\model{NYU}{}$. While the pretrained model was originally trained using input standardization, in this experiment, we evaluated it using images whose pixel values were normalized to the range [0, 1].  As expected, changed the input distribution by compressing its range and altering its mean and standard deviation, which resulted in substantial performance degradation across all three datasets (see $\model{NYU}{'tr}$). Notably, the model reached only 0.56 PR-AUC on the CSAW dataset, reflecting a 37\% drop in performance. However, when test-time statistics were incorporated ($\model{NYU}{'tt*}$), the model's performance substantially recovered, reaching a PR-AUC of 0.89 on CSAW, only 4\% below its best result. Similarly, performance on the HCTP and VinDr datasets were restored to levels comparable with $\model{NYU}{tt*}$. 

These findings suggest that although altering the input distribution leads to performance degradation, this decline is not due to a failure in the convolutional layers’ ability to extract meaningful features. Instead, it is the BN statistics learned during training that result in a divergence from the expected feature distribution, which can be mitigated through test-time adaptation.

Saliency maps in \autoref{results:nyu_tr_tt_smaps}, generated for a CSAW sample, further support this conclusion at a higher level. While the $\model{NYU}{tr}$ model accurately identified the relevant regions, the model evaluated on the altered input distribution ($\model{NYU}{'tr}$) failed to highlight the correct areas. Nevertheless, when test-time BN statistics were applied to the altered input ($\model{NYU}{'tt}$), the model's ability to localize the correct regions was restored.

To better understand the observed gains and limitations of the experiment on the normalized input distribution, we investigated how the BN statistics at test time altered the activation distributions across the layers of the $\model{NYU}{'tr}$ and $\model{NYU}{'tt}$ models. To this end, we generated a mini-batch of 16 images from the CSAW dataset and captured the output distributions of BN layers. We then computed the JS divergence for each layer to quantify the distributional shift. This analysis revealed that the most significant distributional shifts occurred in the early layers of the model (see \autoref{results:nyu_tr_tt_jsdiv}), consistent with the role of early BN layers in normalizing low-level features most affected by domain shift. Additional investigation using kernel density estimations (KDEs) (\ref{appendix:kde:tr_tt_da}) showed that the training-time BN statistics obtained on the NYU dataset led to over-normalized activations and substantially reduced variance in early layers, likely contributing to information loss. In contrast, the use of test-time statistics maintained greater feature variance in early layers and resulted in more stable activation patterns throughout the network, effectively preserving the representational capacity of the data, akin to that observed under standardized inputs ($\model{NYU}{tr}$). Furthermore, while the use of training-time statistics led to increased variance in deeper network layers, test-time statistics promoted a smoother and more stable flow of information throughout the model.

Finally, we assessed the calibration performance of the $\model{NYU}{tr}$ model. As shown in \autoref{fig:calibration_csaw}, despite achieving high PR-AUC scores and strong ranking ability in distinguishing malignant from negative samples, the predicted probabilities were highly underconfident and not well calibrated. Additionally, the reliability diagrams in \autoref{fig:reliability_diagrams} indicate that the model's predictions on the HCTP and VinDr datasets were not only less accurate but also confined to a narrower probability range.

\subsection{Fine-tuning with the HCTP and VinDr Datasets}
In the third experiment, we extended our experiments to the HCTP, VinDr, and HCTP+VinDr datasets. We fine-tuned the pretrained model on these datasets and evaluated it on them. The results are presented in the \autoref{results:table:standardization_malign} with the lines of $\model{HCTP}{}$, $\model{VinDr}{}$, and $\model{HCTP+VinDr}{}$. 

Following fine-tuning on the HCTP dataset, the model's performance improved significantly on the HCTP test set, as expected (see the $\expsetup{HCTP}{HCTP}{tr}$ row in \autoref{results:table:standardization_malign}). Specifically, the PR-AUC increased from 0.60 (without fine-tuning) to 0.86 after fine-tuning. The model also exhibited a notable performance gain on the VinDr dataset ($\expsetup{HCTP}{VinDr}{tr}$). However, on the CSAW dataset, the fine-tuned model ($\expsetup{HCTP}{CSAW}{tr}$) showed a performance drop, with the PR-AUC decreasing from 0.93 to 0.61. These findings align with the observations in \autoref{fig:dataset_hists}, suggesting that the model performs better on test data with pixel intensity distributions similar to those seen during training. The decline in performance on CSAW likely stems from fine-tuning on a distant domain, leading to catastrophic forgetting \citep{karani2018lifelong, french1999catastrophic}, where the model potentially loses information relevant to previously learned domains due to overfitting to the new domain.

\begin{table}[!b]
\caption{Comparison of the best-performing models evaluated on the HCTP dataset. Performance metrics were computed using the decision threshold that maximized the F1 score for each experimental setup (Supplementary Material 1, Table 1 for complete results).}
  \centering
  \bgroup
  \def\arraystretch{1.5}
  \resizebox{\columnwidth}{!}{%
  \begin{tabular}{llccc}
        \cline{2-5}
              & \textbf{Class} & \textbf{Sensitivity} & \textbf{Specificity} &  \textbf{F1-Score} \\
        \hline
        \multirow{2}{*}{$\bm{\model{HCTP}{tr}}$}                 & Benign & 0.943 & 0.443 & 0.755 \\
                                                                 & Malign & 0.841 & 0.761 & 0.809 \\
        \hline
        \multirow{2}{*}{$\bm{\model{HCTP(BNFC)}{tr}}$}           & Benign & 0.943 & 0.420 & 0.748 \\
                                                                 & Malign & 0.807 & 0.727 & 0.776 \\
        \hline
        \multirow{2}{*}{$\bm{\model{HCTP+VinDr(DA\_BNFC)}{tr}}$} & Benign & 0.989 & 0.352 & 0.750 \\
                                                                 & Malign & 0.761 & 0.659 & 0.724 \\
    \hline
    \end{tabular}%
  }
  \egroup
\label{results:table:hctp_ssf}
\end{table}

When test-time BN statistics were applied ($\expsetup{HCTP}{HCTP}{tt} $), the model's performance on the HCTP dataset improved compared to the non-fine-tuned model ($\expsetup{NYU}{HCTP}{tt} $), although it remained lower than the performance achieved using training-time statistics. Similarly, performance on the VinDr dataset ($\expsetup{HCTP}{VinDr}{tt}$) declined compared to using training-time statistics ($\expsetup{HCTP}{VinDr}{tr}$). These findings suggest that training-time BN estimates may offer better generalization within the same or similar domains by providing more representative population-level statistics. Conversely, performance on the CSAW dataset ($\expsetup{HCTP}{CSAW}{tt}$) increased 14-15\% using test-time BN statistics compared to $\expsetup{HCTP}{CSAW}{tr}$. This result suggests that the BN statistics in the fine-tuned model may have overfit to the source domain (HCTP), and that restoring batch statistics from the target domain at test time helps recover adaptation to the target domain. These findings support the idea that BN layers play a central role in catastrophic forgetting by embedding domain-specific activation patterns that hinder performance on unseen domains.

\begin{figure*}[tp]
\centering
\begin{subfigure}{0.32\textwidth}
    \centering
    \includegraphics[width=\textwidth]{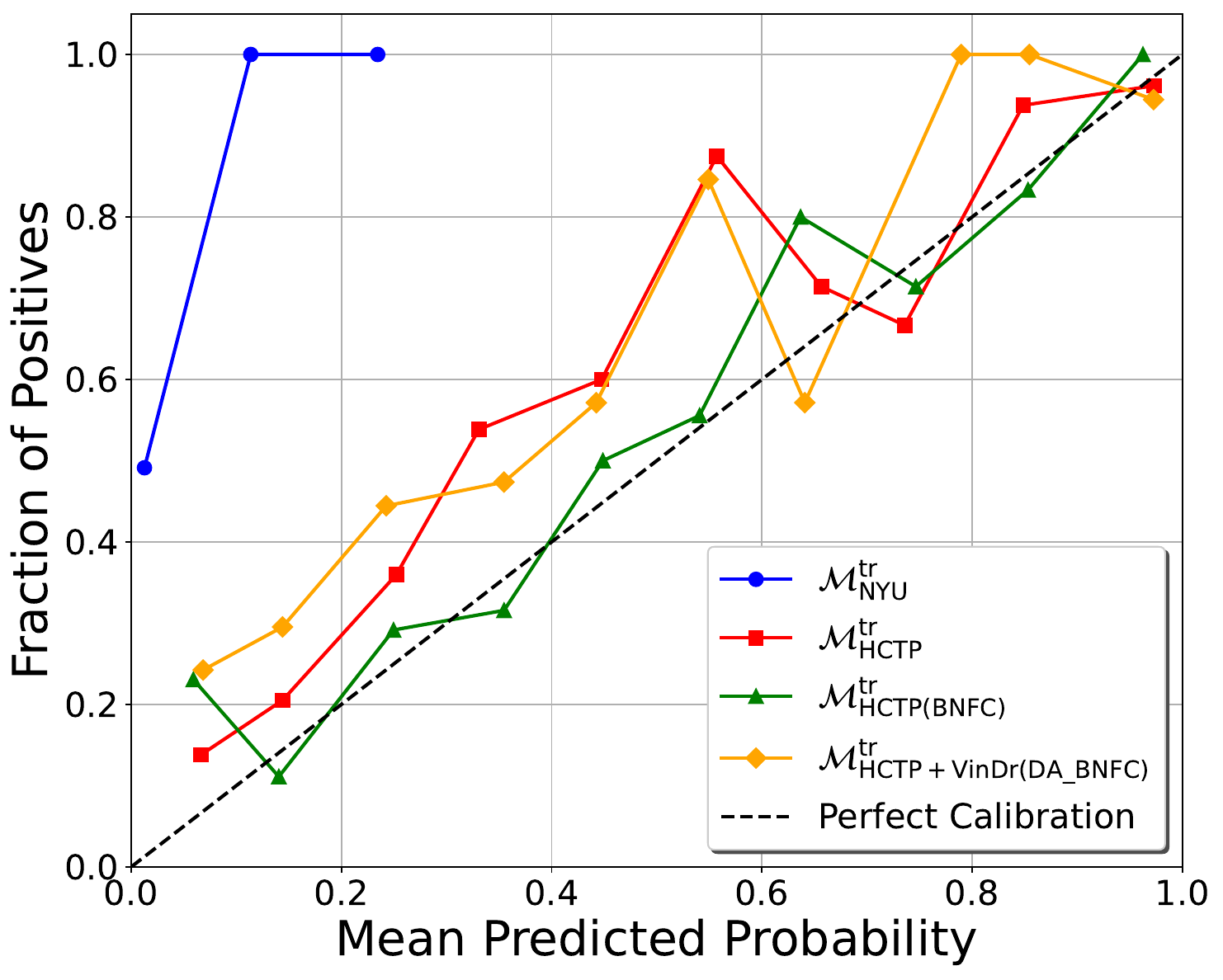}
    \caption{HCTP dataset}
    \label{fig:calibration_hctp}
\end{subfigure}
\hfill
\begin{subfigure}{0.32\textwidth}
    \centering
    \includegraphics[width=\textwidth]{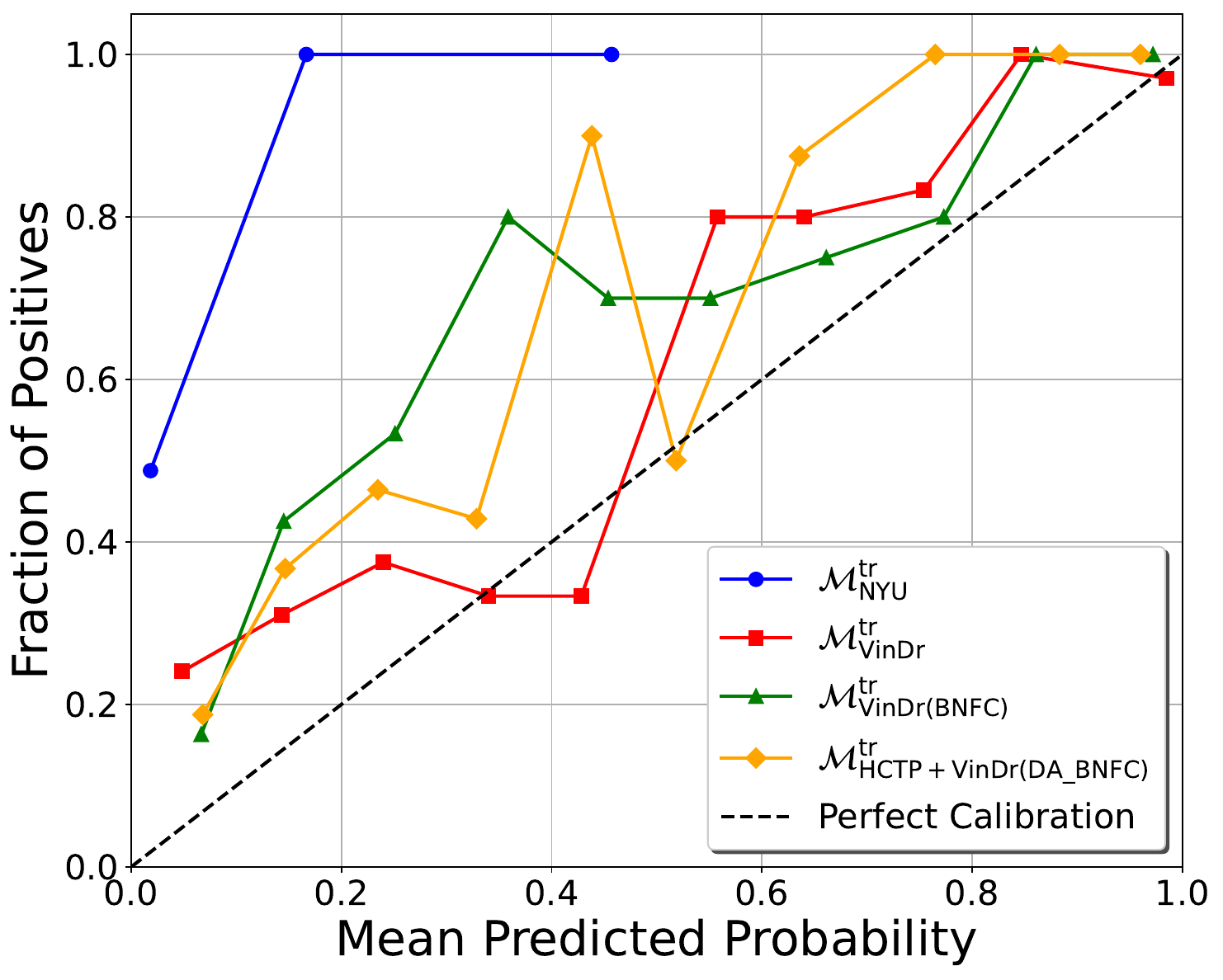}
    \caption{VinDr dataset}
    \label{fig:calibration_vindr}
\end{subfigure}
\hfill
\begin{subfigure}{0.32\textwidth}
    \centering
    \includegraphics[width=\textwidth]{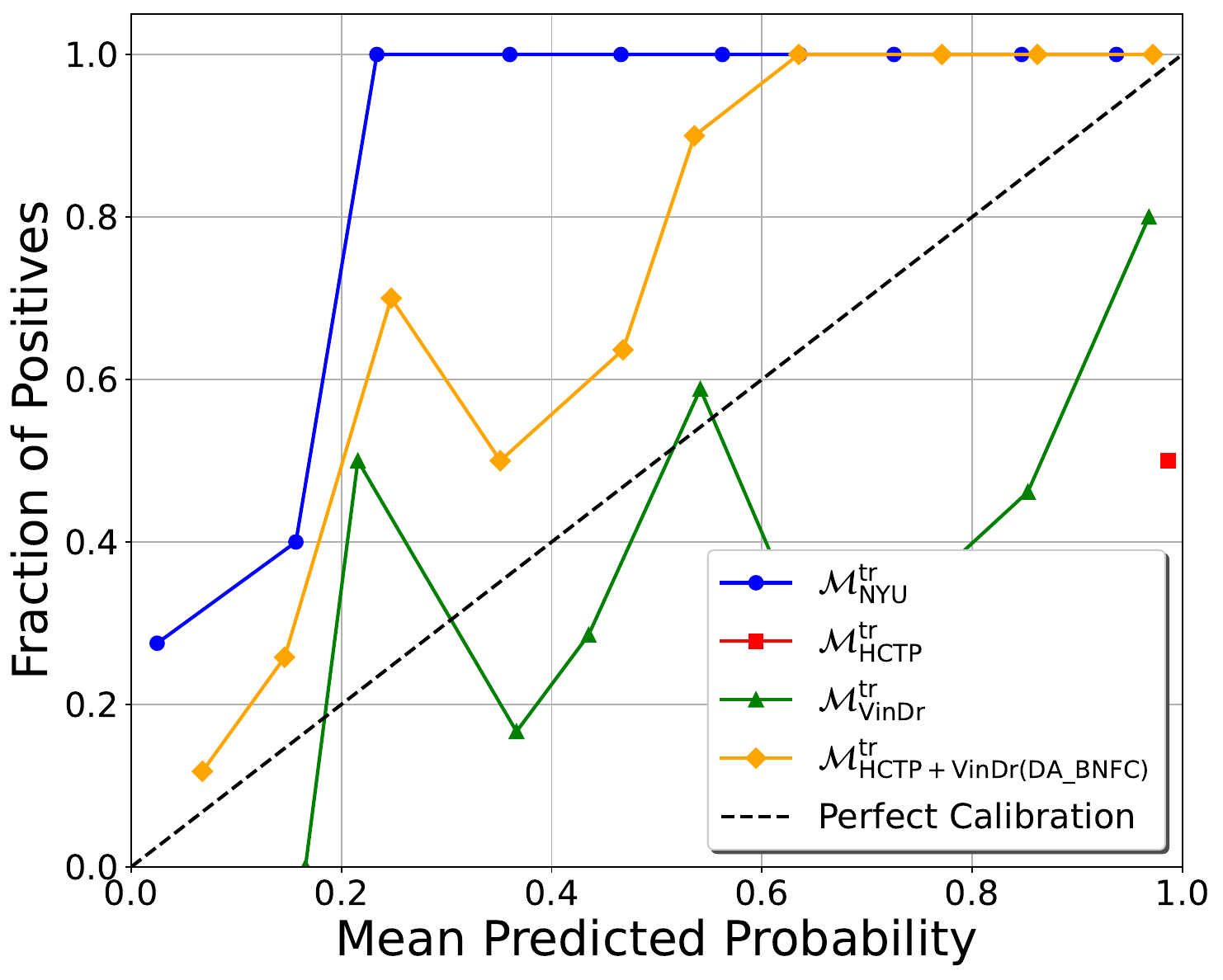}
    \caption{CSAW dataset}
    \label{fig:calibration_csaw}
\end{subfigure}
\caption{Reliability diagrams (probability calibration curves) of the best-performing models evaluated on three independent mammography datasets for malignant class: (a) HCTP, (b) VinDr, and (c) CSAW. For brevity, $\model{HCTP+VinDr(DA_BNFC)}{tr}$ is denoted as $\model{DA_BNFC}{tr}$.}
\label{fig:reliability_diagrams}
\end{figure*}

Fine-tuning on the VinDr dataset exhibited trends similar to those observed with HCTP fine-tuning. The model's performance on the VinDr dataset improved substantially, with the PR-AUC increasing from 0.60 to 0.84 (see the $\expsetup{VinDr}{VinDr}{tr}$ row in \autoref{results:table:standardization_malign}). While the model achieved a 6\% performance gain in PR-AUC on the HCTP dataset ($\expsetup{VinDr}{HCTP}{tr}$) compared to $\expsetup{NYU}{HCTP}{tr}$, it reached a PR-AUC of 0.72 on the CSAW dataset ($\expsetup{VinDr}{CSAW}{tr}$), 11\% higher than when fine-tuned solely on HCTP ($\expsetup{HCTP}{CSAW}{tr}$). 

Although we initially hypothesized that training on a similar domain ($\expsetup{HCTP}{VinDr}{tr}$) would result in greater performance gains, the model trained on the VinDr dataset and evaluated on HCTP ($\expsetup{VinDr}{HCTP}{tr}$) demonstrated only modest improvement, though still a meaningful increase. This may be due to the limited number of training samples in VinDr and the absence of pathological confirmation.  Moreover, incorporating test-time BN statistics did lead to slight improvements in performance on the CSAW dataset, as observed in $\expsetup{VinDr}{CSAW}{tt*}$ and $\expsetup{VinDr}{CSAW}{tt**}$.

To assess whether fine-tuning the model on a larger, combined dataset improves generalizability across domains, we trained the model on the concatenation of the HCTP and VinDr datasets (denoted as $\model{HCTP+VinDr}{}$). This strategy resulted in decrease in PR-AUC on the HCTP and VinDr test sets ($\expsetup{HCTP+VinDr}{HCTP}{tr}$ and $\expsetup{HCTP+VinDr}{VinDr}{tr}$) compared to solely training on the respective dataset. However, it offered more balanced and consistent performance across the two datasets compared to fine-tuning on each dataset separately.

For instance, when the model was fine-tuned only on the VinDr dataset, performance on the HCTP set dropped to 0.66 PR-AUC ($\expsetup{VinDr}{HCTP}{tr}$). In contrast, the joint training setup achieved 0.82 PR-AUC on HCTP ($\expsetup{HCTP+VinDr}{HCTP}{tr}$) while maintaining strong performance on VinDr — highlighting its better cross-domain stability.

However, compared to $\model{NYU}{tr}$, the model’s performance on the CSAW dataset ($\expsetup{HCTP+VinDr}{CSAW}{tr}$) declined more substantially, with PR-AUC dropping to 0.63. This decline may be attributed to catastrophic forgetting. Nevertheless, incorporating test-time statistics mitigated part of this loss, increasing the PR-AUC to 0.79 and 0.78.

In addition to these observations, the reliability diagrams of $\model{HCTP}{tr}$ and $\model{VinDr}{tr}$ (\autoref{fig:calibration_hctp} and \autoref{fig:calibration_vindr}) indicate mild deviations from perfect calibration. Both models exhibit slightly underconfident predictions. Nevertheless, their calibration remains reasonable and could be marginally improved through post-hoc techniques.

Overall, these results suggest that while fine-tuning on the same domain significantly boosts performance within that domain, it can lead to the loss of relevant information for distant domains. Furthermore, fine-tuning provides better population estimates, allowing the model to perform better with training-time statistics than with test-time statistics within the same domain.

\begin{table*}[tp]
  \caption{Models are denoted using the notation defined in ~\autoref{section:experiments}. The model denoted with an apostrophe ($'$) indicates evaluation conducted on input data that has been normalized to the [0, 1] range. For the NYU evaluations, the results were obtained by averaging predictions from the five pretrained models (see footnote~\ref{link:gmic}).}
  
  \small
  \centering
  \bgroup
  \def\arraystretch{1.5}
  \resizebox{\textwidth}{!}{%
  \begin{tabular}{lccccccc}
      \cline{2-7}
      \multirow{2}{*}{~}    
                \cellcolor{white}               &\multicolumn{2}{c} {\textbf{HCTP}}              & \multicolumn{2}{c}{\textbf{VinDr}}            & \multicolumn{2}{c}{\textbf{CSAW}}             \\ \cline{2-7}
                \cellcolor{white}               & \textbf{ROC AUC}      & \textbf{PR AUC}        & \textbf{ROC AUC}      & \textbf{PR AUC}       & \textbf{ROC AUC}      & \textbf{PR AUC}       \\
                        \hline      
      $\bm{\model{NYU}{tr}}$                    & 0.52 ± 0.05           & 0.60 ± 0.04            & 0.55 ± 0.03           & 0.60 ± 0.04           & 0.91 ± 0.00           & 0.93 ± 0.00           \\
      $\bm{\model{NYU}{tt}*}$                   & 0.58 ± 0.01           & 0.63 ± 0.01            & 0.68 ± 0.03           & 0.73 ± 0.03           & 0.87 ± 0.01           & 0.89 ± 0.01           \\
      $\bm{\model{NYU}{tt}**}$                  & 0.57 ± 0.01           & 0.64 ± 0.00            & 0.69 ± 0.03           & 0.74 ± 0.03           & 0.87 ± 0.01           & 0.89 ± 0.01           \\ \hline
      $\bm{\model{NYU}{'tr}}$                   & 0.47 ± 0.05           & 0.49 ± 0.04            & 0.50 ± 0.05           & 0.53 ± 0.04           & 0.53 ± 0.07           & 0.56 ± 0.04           \\
      $\bm{\model{NYU}{'tt} *}$                 & 0.55 ± 0.03           & 0.61 ± 0.02            & 0.67 ± 0.02           & 0.71 ± 0.02           & 0.86 ± 0.02           & 0.89 ± 0.01           \\
      $\bm{\model{NYU}{'tt} **}$                & 0.56 ± 0.04           & 0.62 ± 0.02            & 0.67 ± 0.05           & 0.72 ± 0.04           & 0.86 ± 0.02           & 0.89 ± 0.01           \\ \hline
      $\bm{\model{HCTP}{tr}}$                   & 0.83 ± 0.02           & 0.86 ± 0.02            & 0.70 ± 0.02           & 0.73 ± 0.03           & 0.59 ± 0.07           & 0.61 ± 0.08           \\
      $\bm{\model{HCTP}{tt}*}$                  & 0.71 ± 0.03           & 0.77 ± 0.03            & 0.68 ± 0.03           & 0.71 ± 0.02           & 0.73 ± 0.05           & 0.76 ± 0.04           \\
      $\bm{\model{HCTP}{tt}**}$                 & 0.71 ± 0.03           & 0.76 ± 0.03            & 0.68 ± 0.03           & 0.71 ± 0.02           & 0.73 ± 0.06           & 0.75 ± 0.05           \\ \hline 
      $\bm{\model{VinDr}{tr}}$                  & 0.61 ± 0.04           & 0.66 ± 0.03            & 0.80 ± 0.02           & 0.84 ± 0.02           & 0.67 ± 0.02           & 0.72 ± 0.03           \\
      $\bm{\model{VinDr}{tt}*}$                 & 0.58 ± 0.01           & 0.63 ± 0.01            & 0.67 ± 0.04           & 0.73 ± 0.02           & 0.70 ± 0.03           & 0.74 ± 0.02           \\
      $\bm{\model{VinDr}{tt}**}$                & 0.59 ± 0.01           & 0.64 ± 0.01            & 0.67 ± 0.04           & 0.73 ± 0.02           & 0.71 ± 0.04           & 0.74 ± 0.03           \\ \hline
      $\bm{\model{HCTP+VinDr}{tr}}$             & 0.79 ± 0.02           & 0.82 ± 0.02            & 0.78 ± 0.01           & 0.82 ± 0.01           & 0.61 ± 0.07           & 0.63 ± 0.08           \\
      $\bm{\model{HCTP+VinDr}{tt} *}$           & 0.69 ± 0.01           & 0.76 ± 0.01            & 0.70 ± 0.03           & 0.75 ± 0.02           & 0.74 ± 0.02           & 0.79 ± 0.01           \\
      $\bm{\model{HCTP+VinDr}{tt} **}$          & 0.70 ± 0.01           & 0.75 ± 0.01            & 0.72 ± 0.02           & 0.76 ± 0.01           & 0.75 ± 0.02           & 0.78 ± 0.03           \\ \hline
      $\bm{\model{HCTP(BNFC)}{tr}}$             & 0.82 ± 0.02           & 0.85 ± 0.01            & 0.74 ± 0.07           & 0.78 ± 0.07           & 0.60 ± 0.03           & 0.64 ± 0.01           \\
      $\bm{\model{HCTP(BNFC)}{tr} *}$           & 0.64 ± 0.00           & 0.69 ± 0.01            & 0.64 ± 0.00           & 0.66 ± 0.00           & 0.75 ± 0.01           & 0.79 ± 0.02           \\
      $\bm{\model{HCTP(BNFC)}{tr} **}$          & 0.66 ± 0.01           & 0.70 ± 0.00            & 0.64 ± 0.00           & 0.66 ± 0.00           & 0.81 ± 0.02           & 0.83 ± 0.01           \\ \hline
      $\bm{\model{VinDr(BNFC)}{tr}}$            & 0.66 ± 0.00           & 0.70 ± 0.01            & 0.80 ± 0.01           & 0.83 ± 0.00           & 0.65 ± 0.03           & 0.72 ± 0.02           \\
      $\bm{\model{VinDr(BNFC)}{tr} *}$          & 0.60 ± 0.01           & 0.64 ± 0.01            & 0.65 ± 0.00           & 0.70 ± 0.00           & 0.74 ± 0.01           & 0.79 ± 0.01           \\
      $\bm{\model{VinDr(BNFC)}{tr} **}$         & 0.61 ± 0.01           & 0.67 ± 0.01            & 0.66 ± 0.00           & 0.70 ± 0.01           & 0.78 ± 0.00           & 0.81 ± 0.00           \\ \hline
      $\bm{\model{HCTP+VinDr(BNFC)}{tr}}$       & 0.77 ± 0.01           & 0.81 ± 0.01            & 0.79 ± 0.01           & 0.82 ± 0.00           & 0.64 ± 0.02           & 0.69 ± 0.02           \\
      $\bm{\model{HCTP+VinDr(BNFC)}{tt} *}$     & 0.64 ± 0.00           & 0.69 ± 0.00            & 0.64 ± 0.01           & 0.69 ± 0.01           & 0.69 ± 0.01           & 0.76 ± 0.00           \\
      $\bm{\model{HCTP+VinDr(BNFC)}{tt} **}$    & 0.64 ± 0.00           & 0.68 ± 0.01            & 0.64 ± 0.01           & 0.69 ± 0.01           & 0.76 ± 0.01           & 0.80 ± 0.01           \\ \hline
      $\bm{\model{HCTP+VinDr(DA)}{tr}}$         & 0.82 ± 0.02           & 0.85 ± 0.01            & 0.77 ± 0.03           & 0.82 ± 0.01           & 0.73 ± 0.03           & 0.75 ± 0.03           \\ \hline
      $\bm{\model{HCTP+VinDr(DA\_BNFC)}{tr}}$   & 0.77 ± 0.02           & 0.80 ± 0.01            & 0.79 ± 0.01           & 0.82 ± 0.00           & 0.79 ± 0.02           & 0.82 ± 0.02           \\ \hline
  \end{tabular}%
  }
  \egroup
  \label{results:table:standardization_malign}
\end{table*}

\subsection{DoSReMC: Fine-Tuning Only BN and FC Layers}
Previous experiments showed that even without fine-tuning, a model's performance can be increased using test-time BN statistics under domain shift. This is achieved by updating first- and second-order statistics of the model with respect to test batches. However, the BN layer’s scaling ($\gamma$) and shifting ($\beta$) parameters adjust the effect of these statistics and regulate their influence on the layer’s output. \citep{karani2018lifelong} showed that adapting BN layers increased segmentation performance on brain MRI segmentation. \citep{frankle2020training} explored the expressive power of these parameters on ImageNet and CIFAR datasets. They demonstrated that tuning only the BN and output layers, even with randomly initialized convolutional weights, can substantially improve performance. While their focus was on understanding BN’s contribution to a model’s representational capacity, we extend this idea to domain adaptation for mammography classification. To investigate whether tuning only the BN and FC layers improves performance under distribution shift, we conducted experiments restricting training to these components.

The $\model{HCTP(BNFC)}{tr}$, $\model{VinDr(BNFC)}{tr}$, and $\model{HCTP+VinDr(BNFC)}{tr}$ rows in \autoref{results:table:standardization_malign} present the results obtained by DoSReMC, which fine-tunes only the BN and FC layers on the HCTP, VinDr, and HCTP+VinDr datasets. This strategy demonstrates that such selective adaptation can significantly improve performance, aligning with the observations of \citep{karani2018lifelong,frankle2020training}.

The DoSReMC model achieved performance comparable to the fully fine-tuned models of $\model{HCTP(BNFC)}{tr}$ and $\model{VinDr(BNFC)}{tr}$, with results only about 1\% lower than the upper bounds. Compared to $\expsetup{HCTP}{HCTP}{tr}$, the ROC-AUC of $\expsetup{HCTP(BNFC)}{HCTP}{tr}$ decreased by 0.01. The ROC-AUC of $\model{VinDr(BNFC)}{tr}$ remained unchanged at 0.80. While $\model{HCTP+VinDr(BNFC)}{tr}$ yielded close performance to $\model{HCTP+VinDr}{tr}$ on the HCTP and VinDr datasets, it reached higher ROC- and PR-AUC scores compared to $\expsetup{HCTP+VinDr}{CSAW}{tr}$.

\begin{figure}[t]
\centering
  \begin{subfigure}{0.16\textwidth}
    \includegraphics[width=\textwidth, height=5cm]{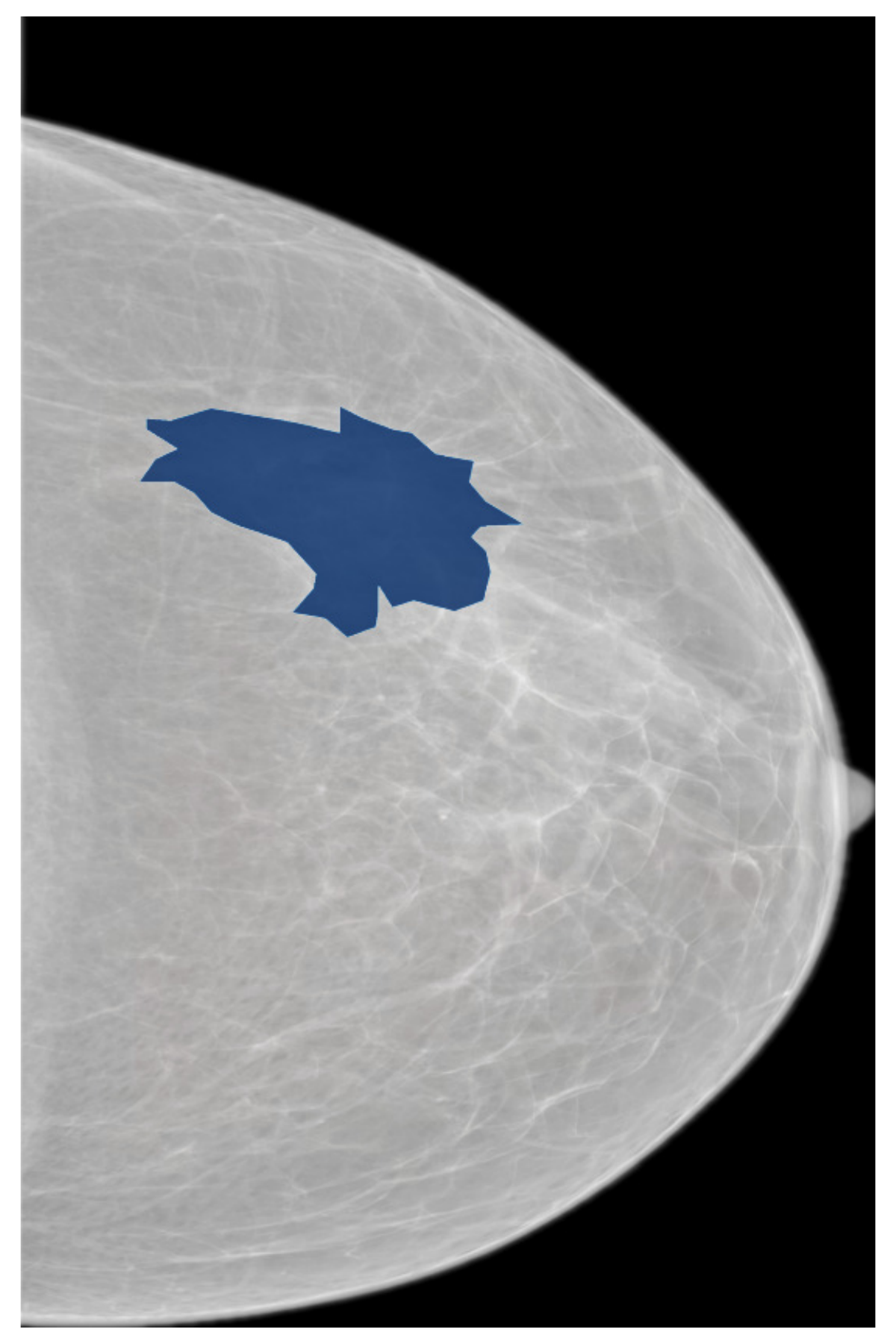}
    \caption{$\model{NYU}{tr}$}
  \end{subfigure}%
  \begin{subfigure}{0.16\textwidth}
    \includegraphics[width=\textwidth, height=5cm]{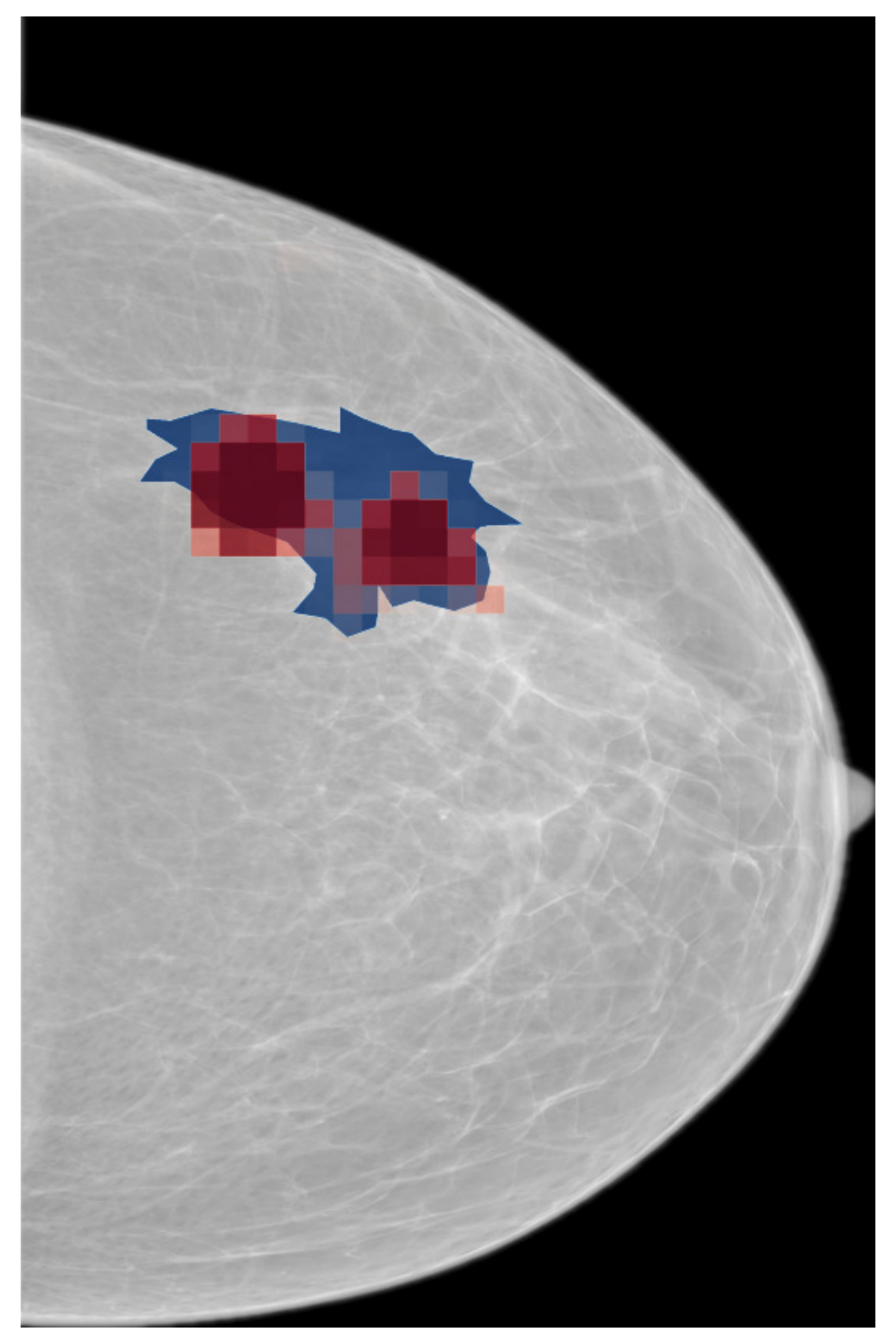}
    \caption{$\model{HCTP}{tr}$} 
  \end{subfigure}%
  \begin{subfigure}{0.16\textwidth}
    \includegraphics[width=\textwidth, height=5cm]{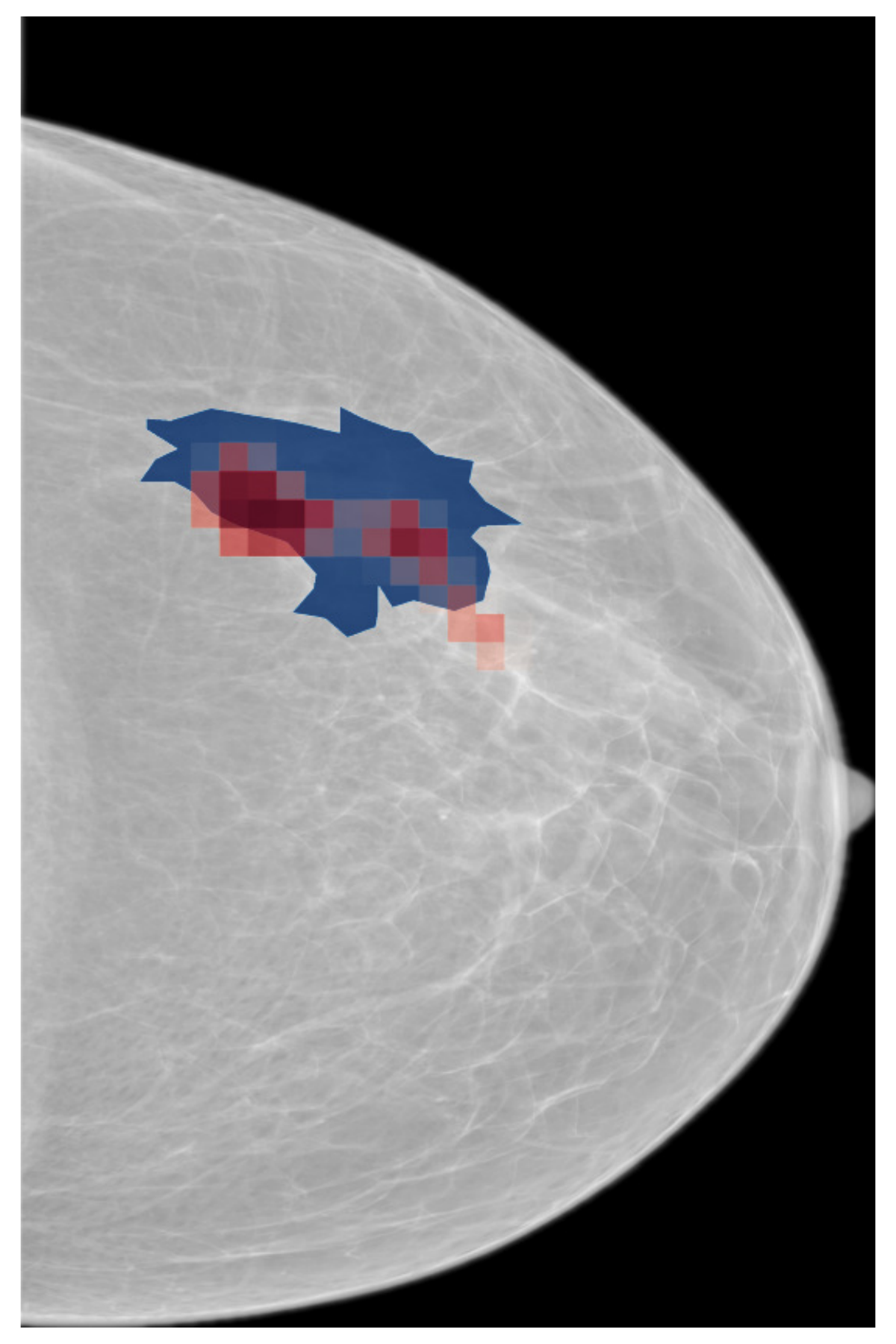}
    \caption{$\model{HCTP(BNFC)}{tr}$} 
  \end{subfigure}%
\caption{Saliency maps generated by $\model{NYU}{tr}$, $\model{HCTP}{tr}$, $\model{HCTP(BNFC)}{tr}$  models a sample from the HCTP dataset. Blue regions represent the ground truth annotations, while red regions highlight the saliency maps corresponding to the malignant class.}
\label{results:hctp_bnfc_smaps}
\end{figure}

In addition, we evaluated the $\model{HCTP(BNFC)}{tr}$ model on the HCTP dataset using a threshold that maximized the F1 score (see \autoref{results:table:hctp_ssf}). The results indicate that DoSReMC achieved performance comparable to the fully fine-tuned model ($\model{HCTP}{tr}$). For the malignant class, its sensitivity and specificity scores were lower by less than 0.04 compared to $\model{HCTP}{tr}$. For the benign class, the achieved sensitivity score was identical to that of $\model{HCTP}{tr}$, and its F1 score was only 0.01 lower. These results demonstrate that $\model{HCTP(BNFC)}{tr}$ maintains strong malignant-class discrimination while achieving efficient calibration, although its lower specificity for benign cases suggests a tendency to classify negative samples as benign. Also, as shown in \autoref{fig:calibration_hctp}, $\model{HCTP(BNFC)}{tr}$ achieved better calibration, whereas $\model{VinDr(BNFC)}{tr}$ produced more underconfident predictions compared to $\model{VinDr}{tr}$.

The saliency map in \autoref{results:hctp_bnfc_smaps} further supports these findings, showing that $\expsetup{HCTP(BNFC)}{HCTP}{tr}$ highlights similar regions of interest as the fully fine-tuned model ($\model{HCTP}{tr}$).

Moreover, DoSReMC demonstrated improved performance on similar domains. For instance, when the model was fine-tuned on the HCTP dataset, its PR-AUC score on the VinDr dataset was 0.73 ($\expsetup{HCTP}{VinDr}{tr}$), whereas $\expsetup{HCTP(BNFC)}{VinDr}{tr}$ achieved 0.78. We hypothesize that convolutional filters learned from a large dataset such as NYU are more representative, and that full fine-tuning may disrupt these kernels. Restricting fine-tuning to the BN and FC layers therefore leads to better generalization. 

\begin{figure*}[t]
\centering
    \includegraphics[width=14cm, height=6cm]{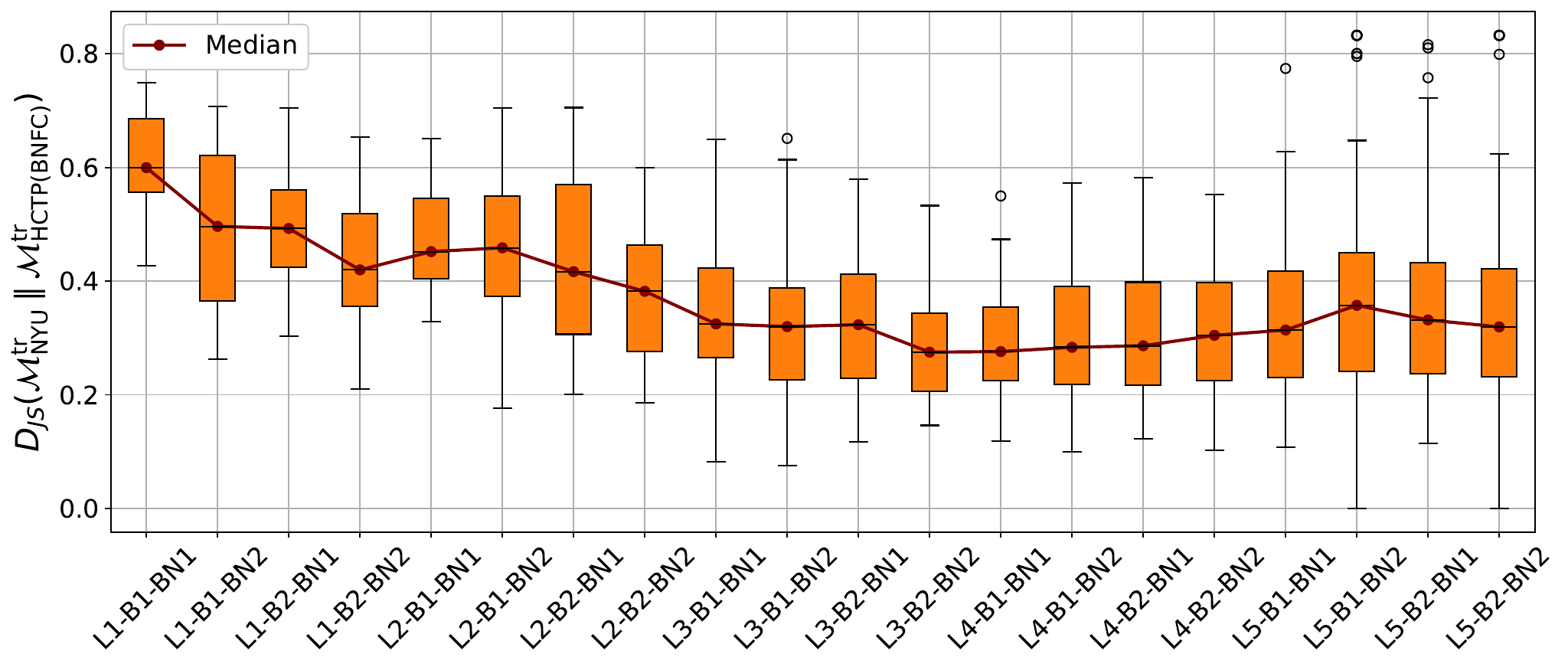}
\caption{JS divergence across BN layers in the Global Module (ResNet-22), using $\model{NYU}{tr}$ versus $\model{HCTP(BNFC)}{tr}$ on a sample batch from HCTP.}
\label{results:nyu_bnfc_jsdiv}
\end{figure*}

Further analysis of the BN output distributions of the $\model{HCTP(BNFC)}{tr}$ supports these findings. As observed previously on the CSAW dataset, the most significant divergence occurred in the early layers of the global model compared to the results obtained with the pretrained model, while the deeper layers exhibited smaller shifts (\autoref{results:nyu_bnfc_jsdiv}). Additionally, our KDE analysis revealed that fine-tuning only the BN and FC layers enables the model to capture more representative input features in the early layers, as reflected by broader and more dispersed BN activation distributions (\ref{appendix:kde:nyu_bnfc_hctp}). These results suggest that adapting the moving averages and learnable parameters of early BN layers to the target distribution may be a key factor in achieving performance gains and reducing feature collapse.

\subsection{DoSReMC with Domain-Adversarial Training}
In this experiment, we evaluated whether integrating DAT into DoSReMC could further improve cross-domain generalization. The $\model{HCTP+VinDr(DA\_BNFC)}{tr}$ rows in \autoref{results:table:standardization_malign} present the results obtained through partial DAT on HCTP+VinDr. In the fourth experiment, we showed that fine-tuning only the BN and FC layers (DoSReMC) can achieve performance comparable to fully trained models. We attribute this to the fact that the pretrained $\model{NYU}{}$ model's convolutional layers learned robust feature representations from a large dataset, and that fine-tuning only the BN and FC layers allows targeted adaptation on smaller datasets. Although this approach provides cost-efficient fine-tuning for domain adaptation, its generalizability to unseen domains remains limited (see $\expsetup{HCTP+VinDr(BNFC)}{CSAW}{tr}$ in \autoref{results:table:standardization_malign}).

By incorporating  DAT into DoSReMC, the $\model{HCTP+VinDr(DA\_BNFC)}{tr}$ model achieved the most consistent performance across all three datasets using a single model. While its performance on the HCTP dataset ($\expsetup{HCTP+VinDr(DA\_BNFC)}{HCTP}{tr}$) was lower than in the $\expsetup{HCTP}{HCTP}{tr}$ experiment, it still achieved a ROC-AUC of 0.77 and a PR-AUC of 0.80, slightly lower than that of $\model{HCTP+VinDr(BNFC)}{tr}$. On the CSAW dataset, it outperformed all other fine-tuned models, reaching 0.79 ROC-AUC and 0.82 PR-AUC.

We further compare $\model{HCTP+VinDr(DA\_BNFC)}{tr}$ against an established domain adaptation method, DAT, implemented by applying adversarial adaptation to all layers ($\model{HCTP+VinDr(DA)}{tr}$), as shown in \autoref{results:table:standardization_malign}. Although both approaches yielded comparable results, restricting fine-tuning to the BN and FC layers was computationally more efficient (see \autoref{results:table:forward_backward_times}). By keeping convolutional parameters frozen, this configuration achieved nearly 10 times faster gradient updates and approximately 20\% lower memory consumption (see \autoref{results:table:forward_backward_times}).

\begin{table}[t]
\caption{Comparison of the best-performing models evaluated on the CSAW dataset. Due to label inconsistency in negative/benign cases in the dataset, performance metrics for benign class were excluded. The metrics were computed using the decision threshold that maximized the F1 score for each experimental setup (Supplementary Material 1, Table 2 for complete results).}

  \centering
  \bgroup
  \def\arraystretch{1.5}
  \resizebox{\columnwidth}{!}{%
  \begin{tabular}{lccccc}
        \cline{2-4}
        & \textbf{Sensitivity} & \textbf{Specificity} & \textbf{F1-Score} \\
        \hline
        $\bm{\model{NYU}{tr}}$                   & 0.925 & 0.736 & 0.845 \\    \hline
        $\bm{\model{HCTP}{tr}}$                  & 0.868 & 0.396 & 0.702 \\    \hline
        $\bm{\model{VinDr}{tr}}$                 & 0.925 & 0.245 & 0.690 \\    \hline
        $\bm{\model{HCTP+VinDr}{tr}}$            & 0.925 & 0.226 & 0.685 \\    \hline
        $\bm{\model{HCTP+VinDr(DA\_BNFC)}{tr}}$  & 0.868 & 0.660 & 0.786 \\    \hline
    \end{tabular}%
  }
  \egroup
\label{results:table:csaw_ssf}
\end{table}

Moreover, this approach achieved the highest specificity and F1 scores among all fine-tuned models, as shown in \autoref{results:table:csaw_ssf}. While other models reached higher sensitivity, their lower specificity and F1 scores indicate a higher tendency toward false positives. Furthermore, the reliability curves of $\model{HCTP+VinDr(DA\_BNFC)}{tr}$ demonstrate better calibration compared to $\model{NYU}{tr}$, indicating more reliable confidence estimates, as shown in \autoref{fig:calibration_csaw}. In addition, the prediction probabilities of $\model{HCTP}{tr}$ on the CSAW dataset were highly saturated near 1, explaining its limited discriminative ability and the resulting PR-AUC score of 0.61 (see \autoref{results:table:standardization_malign}).

\begin{table*}[t]
\caption{Forward and backward pass times and memory consumption measured on RTX 3090. All times are reported in milliseconds, and memory usage in megabytes (MB).}
    
  \small
  \centering
  \bgroup
  \def\arraystretch{1.5}
  \resizebox{0.75\textwidth}{!}{%
  \begin{tabular}{lcccc}
      \toprule
      \textbf{Model} & \textbf{Batch Size} & \textbf{Forward (ms)} & \textbf{Backward (ms)} & \textbf{Memory Consumption (MB)} \\
      \midrule
      $\bm{\model{HCTP+VinDr(DA)}{tr}}$       & \multirow{2}{*}{16}  & \multirow{2}{*}{$\sim$121.2} & $\sim$384.7 & 12065 \\
      $\bm{\model{HCTP+VinDr(DA\_BNFC)}{tr}}$ &  &  & $\sim$37.1 & 9758 \\
      \bottomrule
  \end{tabular}%
  }
  \egroup
  \label{results:table:forward_backward_times}
\end{table*}

These findings reinforce the conclusion that BN layers are highly sensitive to domain shifts, consistent with our previous experiments. DoSReMC, especially when combined with DAT, proves to be an effective strategy for mitigating such shifts and improving model generalizability, while also enabling faster and more computationally efficient gradient updates.

Nonetheless, convolutional layers also contribute meaningfully to in-domain performance. While DoSReMC+DAT improved generalization across domains, it does not fully match the results of full or partial fine-tuning $\model{HCTP(BNFC)}{tr}$ on the HCTP dataset, suggesting that adapting the convolutional filters remains important for maximizing in-domain performance.

\subsection{Ablation Study}
In this section, we present an ablation study to quantify the contribution of individual components of our method under domain shift. We consider three groups of baselines: (i) selective fine-tuning, (ii) lightweight adaptation via alternative normalization strategies, and (iii) simple data-level preprocessing methods such as histogram matching.

\subsubsection{Impact of Selective Fine-Tuning (FC vs.\ BN vs.\ BN+FC vs.\ Full Fine-Tuning)}

\begin{table}[!b]
  \caption{Comparison of selective fine-tuning strategies on the HCTP dataset. $\model{HCTP(FC)}{tr}$ denotes models where only the FC layers are fine-tuned, while $\model{HCTP(BN)}{tr}$ denotes models where only the BN layers are fine-tuned. For clarity and completeness, the $\model{HCTP(BNFC)}{tr}$ and $\model{HCTP}{tr}$ results are reproduced from \autoref{results:table:standardization_malign} to enable direct comparison.}
  \small
  \centering
  \bgroup
  \def\arraystretch{1.5}
  \begin{tabular}{lcc}
  \cline{2-3}
                                      & \textbf{ROC AUC}      & \textbf{PR AUC}     \\  \hline
    $\bm{\model{HCTP(FC)}{tr}}$       & 0.65 ± 0.00           & 0.68 ± 0.01         \\
    $\bm{\model{HCTP(BN)}{tr}}$       & 0.80 ± 0.01           & 0.82 ± 0.01         \\
    $\bm{\model{HCTP(BNFC)}{tr}}$     & 0.82 ± 0.02           & 0.85 ± 0.01         \\
    $\bm{\model{HCTP}{tr}}$           & 0.83 ± 0.02           & 0.86 ± 0.02         \\ \hline

  \end{tabular}%
  \egroup
  \label{results:ablation_fc_bn}
\end{table}

To isolate which model components contribute most to performance changes under domain shift, we fine-tuned only the FC layers and only the BN layers, and compared these settings with joint BN+FC fine-tuning ($\model{HCTP(BNFC)}{tr}$) as well as full fine-tuning of all layers ($\model{HCTP}{tr}$), including the convolutional backbone. The results are summarized in \autoref{results:ablation_fc_bn}. Each experiment was repeated four times with different random seeds, and we report the mean and standard deviation across runs. Extended results for the FC-only and BN-only settings are provided in Supplementary Material 3 and Supplementary Material 4, respectively.

This ablation confirms that fine-tuning only the BN and FC layers under domain shift substantially improves performance, whereas fine-tuning FC layers alone provides limited benefit. Moreover, adapting only BN layers already yields strong gains; however, given the minimal additional overhead of fine-tuning FC layers jointly with BN, the BN+FC strategy offers a more balanced and effective solution, supporting the design choice of DoSReMC.

\subsubsection{Comparison with Alternative Normalization Strategies (LayerNorm vs.\ GroupNorm vs.\ BatchNorm)}

\begin{table*}[ht]
  \caption{Comparison of alternative normalization strategies under full fine-tuning. GN and LN models were trained on the HCTP dataset and evaluated across HCTP, VinDr, and CSAW test sets. The NYU-pretrained and BN full fine-tuning results are reproduced from \autoref{results:table:standardization_malign} for reference. $\model{HCTP(BN)}{tr}$ denotes the same BN-based model as $\model{HCTP}{tr}$ in \autoref{results:table:standardization_malign}, and is reintroduced here to avoid ambiguity.}

  \small
  \centering
  \bgroup
  \def\arraystretch{1.5}
  \resizebox{0.7\textwidth}{!}{
  \begin{tabular}{lccccccc}
      \cline{2-7}
      \multirow{2}{*}{~}
                \cellcolor{white}        &\multicolumn{2}{c} {\textbf{HCTP}}    & \multicolumn{2}{c}{\textbf{VinDr}}    & \multicolumn{2}{c}{\textbf{CSAW}}             \\ \cline{2-7}
                \cellcolor{white}        & \textbf{ROC AUC} & \textbf{PR AUC}   & \textbf{ROC AUC} & \textbf{PR AUC}    & \textbf{ROC AUC}      & \textbf{PR AUC}       \\ \hline
      $\bm{\model{NYU}{tr}}$             & 0.52           & 0.60                & 0.55           & 0.60                 & 0.91            & 0.93                        \\
      $\bm{\model{HCTP(LN)}{tr}}$        & 0.74           & 0.78                & 0.63           & 0.66                 & 0.66            & 0.70                        \\
      $\bm{\model{HCTP(GN)}{tr}}$        & 0.78           & 0.82                & 0.70           & 0.74                 & 0.69            & 0.68                        \\
      $\bm{\model{HCTP(BN)}{tr}}$        & 0.83           & 0.86                & 0.70           & 0.73                 & 0.59            & 0.61                        \\ \hline
  \end{tabular}
  }
  \egroup
  \label{results:ablation_layer_norm}
\end{table*}

In this section, we evaluated alternative normalization strategies to assess whether they provide improved robustness to domain shift compared to BN. Specifically, we replaced all BN layers in the pretrained network with Group Normalization (GN) \citep{wu2018group} and Layer Normalization (LN) \citep{ba2016layer}. For GN, the number of groups was set to 8, following prior findings that GN performance is relatively insensitive to the exact number of groups \citep{wu2018group}. One of the available NYU-pretrained models was selected at random as the baseline for this experiment, as the goal was comparative evaluation rather than absolute performance.

These experiments were conducted by fully fine-tuning the resulting models on the HCTP dataset. Since GN and LN do not rely on batch-wise or moving-average statistics, their inference behavior is identical at training and test time; therefore, only training-time results are reported in \autoref{results:ablation_layer_norm}. To limit computational overhead, each normalization variant was trained once using a fixed random seed.

Replacing BN layers with LN or GN disrupts the original pretrained model structure, as the pretrained weights were optimized jointly with BN statistics. Nevertheless, we include this experiment to assess whether normalization strategies that are independent of batch statistics yield improved cross-domain robustness.

The results indicate that LN underperforms both GN and BN on the HCTP and VinDr datasets. GN consistently improves over LN across datasets but generally remains inferior to BN when models are trained on the HCTP. Overall, BN yields the strongest performance among the evaluated normalization strategies, which is consistent with prior findings showing that BN is more effective than GN and LN when sufficient batch statistics are available \citep{wu2018group}.

\subsubsection{Test-Time Histogram Matching}
To assess whether a simple data-level strategy can mitigate performance degradation under domain shift, we evaluated test-time histogram matching (HM) as a baseline input-level adaptation method. Specifically, we matched CSAW test images to the source-domain intensity distributions of HCTP and VinDr and compared the resulting performance to models evaluated using models only BN and FC layers fine-tuned (see \autoref{results:ablation_hm}).

\begin{table}[b]
  \caption{Test-time HM results on the CSAW dataset. For reference, fine-tuning only BN and FC layer results are reproduced from \autoref{results:table:standardization_malign}.}
  \small
  \centering
  \bgroup
  \def\arraystretch{1.5}
  \begin{tabular}{lcc}
  \cline{2-3}
                                    & \textbf{ROC AUC}      & \textbf{PR AUC}     \\ \hline
    $\bm{\model{HCTP(HM)}{tr}}$     & 0.50                  & 0.54          \\
    $\bm{\model{VinDr(HM)}{tr}}$    & 0.65                  & 0.69          \\
    $\bm{\model{HCTP(BNFC)}{tr}}$   & 0.60                  & 0.64          \\
    $\bm{\model{VinDr(BNFC)}{tr}}$  & 0.65                  & 0.72          \\\hline

  \end{tabular}%
  \egroup
  \label{results:ablation_hm}
\end{table}

For this experiment, we computed average reference histograms using images from the HCTP and VinDr test sets, with 4096-bin discretization, and applied histogram matching to each CSAW test image prior to inference. This procedure reflects a deployment-time preprocessing strategy and does not use any label information. To limit computational overhead, we report results from a single trained model for each source domain, as histogram matching did not yield performance gains.

As shown in \autoref{results:ablation_hm}, test-time histogram matching results in degraded performance on the CSAW dataset for both source domains (HCTP and VinDr). These findings indicate that aligning global intensity distributions alone is insufficient to address scanner-induced domain shift and may even distort discriminative image characteristics, highlighting the advantage of feature-level adaptation via BN statistics.


\section{Discussion \& Conclusion}
CNNs are among the most widely used deep learning architectures in medical image analysis, consistently demonstrating strong performance across various disease recognition tasks. However, their effectiveness is highly dependent on the characteristics of the training data, which can significantly hinder their generalizability in real-world applications. In this study, we investigated this dependency from an architectural perspective and conducted a systematic analysis to identify the key factors contributing to these limitations.

To this end, we introduced the HCTP dataset and incorporated the CSAW and VinDr datasets, which consist of mammography images acquired using GE, Hologic, and Siemens devices, respectively. We compared pixel intensity distributions across devices to highlight domain-level differences. We employed a CNN-based deep learning model pretrained on a large dataset comprising images from Hologic and Siemens devices. Evaluation with this pretrained model demonstrated that applying test-time BN statistics on the VinDr dataset significantly improved PR-AUC performance. Furthermore, when a distribution shift was induced in the CSAW dataset, test-time BN statistics effectively restored the model’s performance. To better understand how the moving averages of BN statistics contribute to performance degradation under distributional shifts, we analyzed the data flow through the model’s layers. Our analysis of the input distributions to the global module layers showed that training-time BN statistics compress feature distributions into a narrower range on the normalized CSAW inputs, causing information loss when the target domain differs significantly. In contrast, using test-time BN statistics helped preserve feature variability and recover performance on distributionally altered inputs. The saliency maps, used for visual comparison, provide high-level support for this conclusion. Our further analysis of the real-world distributions in the HCTP, VinDr, and CSAW datasets also aligns with these observations. These findings highlight the critical role of BN statistics in mitigating performance degradation under distributional shifts and demonstrate their effectiveness in preserving model accuracy on out-of-distribution data.

Building on these findings, we propose DoSReMC, a targeted adaptation strategy that preserves pretrained convolutional kernels while adapting BN statistics and parameters for the target domain. By freezing convolutional layers and updating only BN and FC components, DoSReMC achieved substantial performance gains even under domain mismatch. Extending this approach with DAT, applied exclusively to BN and FC layers, yielded the most consistent performance across all three datasets compared to conventional DAT applied to all layers. This selective integration mitigates potential side effects of DAT—such as distortion of well-learned convolutional features—while still enhancing cross-domain generalizability.

In addition to its robustness, DoSReMC also reduce training cost and requires no architectural modifications, making it easily integrable into existing models. Unlike methods that rely on test-time batch statistics—which demand multiple images to ensure stable inference—DoSReMC operates effectively on a single image, enabling efficient deployment in clinical workflows without increasing inference latency.

Nevertheless, while DoSReMC+DAT was particularly effective on the CSAW dataset, it did not surpass the best in-domain fine-tuning results, indicating that further investigation is needed to assess its behavior on additional unseen domains.

The key takeaway from our experiments is the strong dependence of BN layers on the characteristics of the training data. When the test data differs substantially from the training distribution, the BN statistics learned during training can impair model performance, even if the convolutional layers have successfully learned effective feature extractions. This suggests that, while BN facilitates optimization and convergence, it can also amplify performance degradation under domain shift. Although this issue has been addressed in other domains \citep{lim2023ttn, niu2023towards, zhao2023delta}, to the best of our knowledge, this is the first comprehensive study to investigate it in the context of mammography classification.

One potential approach to mitigate this issue involves collecting heterogeneous data from multiple sources to ensure more representative training samples. Nonetheless, the dynamic nature of real-world data may constrain the sustained effectiveness of this approach. Alternatively, fine-tuning only the BN parameters specific to each domain presents a more efficient and feasible strategy. In federated learning scenarios, for instance, exchanging only fine-tuned BN parameters can effectively reduce communication overhead and alleviate privacy concerns.

Another possible direction is to align pixel intensity distributions across datasets. However, simple preprocessing techniques such as histogram matching or histogram equalization have shown limited or inconsistent improvements under domain shifts in prior studies \citep{sendra2022domain}, which is consistent with our findings in \autoref{results:ablation_hm}. In contrast, more advanced augmentation-based strategies tailored for mammography have demonstrated more promising results \citep{lauritzen2023robust}.

Adaptive test-time adaptation (TTA) methods \citep{lim2023ttn, zhao2023delta} present another promising approach to address this issue. As future work, we plan to explore alternative test-time normalization and domain adaptation strategies to systematically compare our approach with existing methods. This will help develop more robust deep learning solutions for mammography classification and to facilitate the deployment of medical image analysis models in real-world clinical settings.

\section*{Ethics statement}
Ethical approval for the use of patient data was granted by the Non-Interventional Clinical Trials Ethics Committee of Hacettepe University Faculty of Medicine (Approval No: 2020/19-54). 

\section*{Acknowledgments}
We gratefully acknowledge our late colleague and friend, Mete Can Kaya, whose contributions and spirit continue to inspire this work. This study is dedicated to his memory. 

This study was supported by the Scientific and Technological Research Council of Turkey (TÜBİTAK) under the 1501 Industrial R\&D Projects Grant Program, Project No: 3200665 (Project Official URL: \url{https://www.icterra.com/project/midas/}). 

Finally, we extend our deepest appreciation to H. Vedat Uslu, General Manager of ICterra Information and Communication Technologies, for his unwavering commitment to our research. 

\clearpage
\bibliographystyle{elsarticle-num}
\bibliography{main_num}

\onecolumn 
\appendix

\clearpage
\section{Kernel Density Estimations}
\subsection{\texorpdfstring{KDEs of BN outputs from $\model{NYU}{tr}$, $\model{NYU}{'tr}$, and $\model{NYU}{'tt}$}{KDEs of BN outputs from NYU-tr, NYU'-tr, and NYU'-tt}}

\label{appendix:kde:tr_tt_da}
\begin{figure}[H]
    \centering
    \begin{subfigure}[t]{0.9\textwidth}
        \centering
        \includegraphics[width=\textwidth, height=5.5cm]{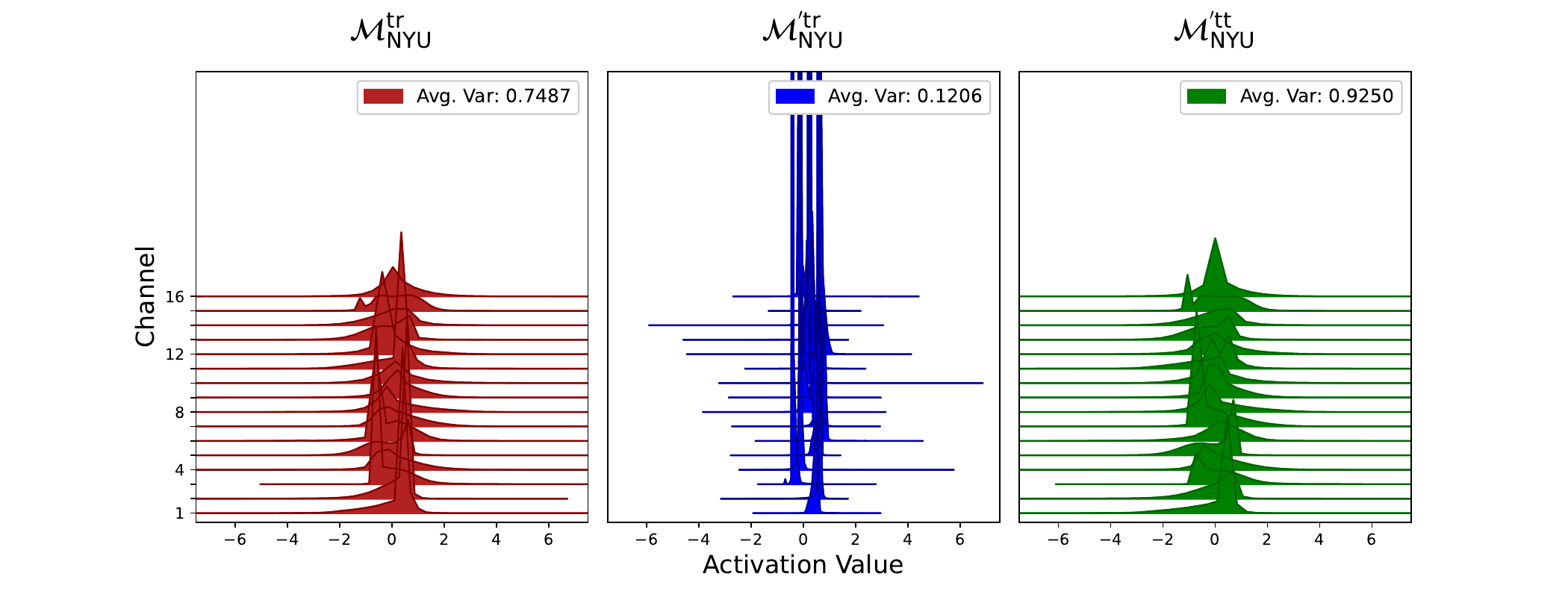}
        \caption{ResNet Layer: 1, Block: 2, BN: 2}
        \label{fig:ap:csaw:l1-b2-bn2}
    \end{subfigure}
    
    \begin{subfigure}[t]{0.9\textwidth}
        \centering
        \includegraphics[width=\textwidth, height=14cm]{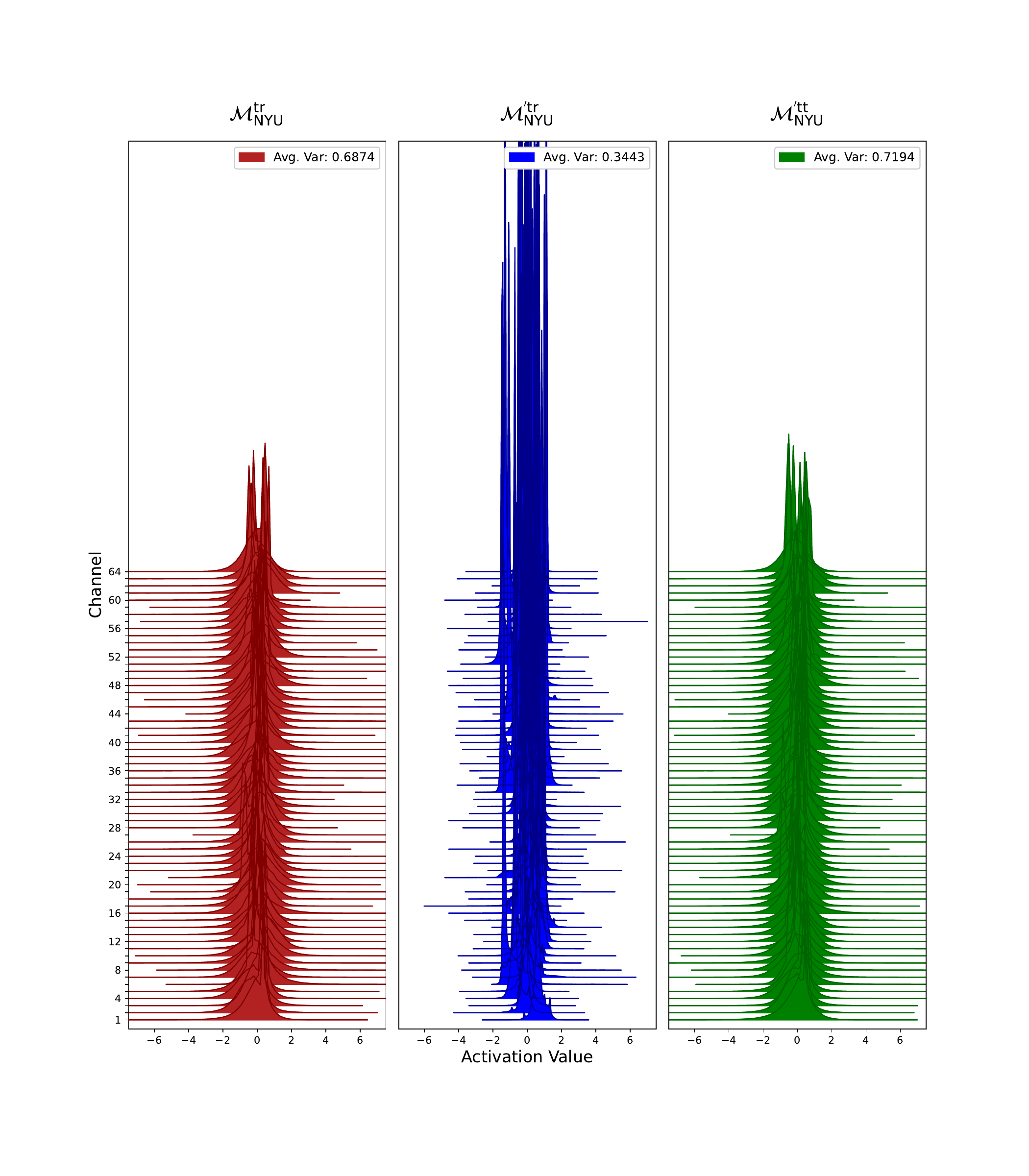}
        \vspace{-15mm}
        \caption{ResNet Layer: 3, Block: 2, BN: 2}
        \label{fig:ap:csaw:l3-b2-bn2}
    \end{subfigure}
    \end{figure}

    \begin{figure}[H]
    \ContinuedFloat
    \centering
    \begin{subfigure}[t]{0.9\textwidth}
        \centering
        \includegraphics[width=\textwidth, height=20cm]{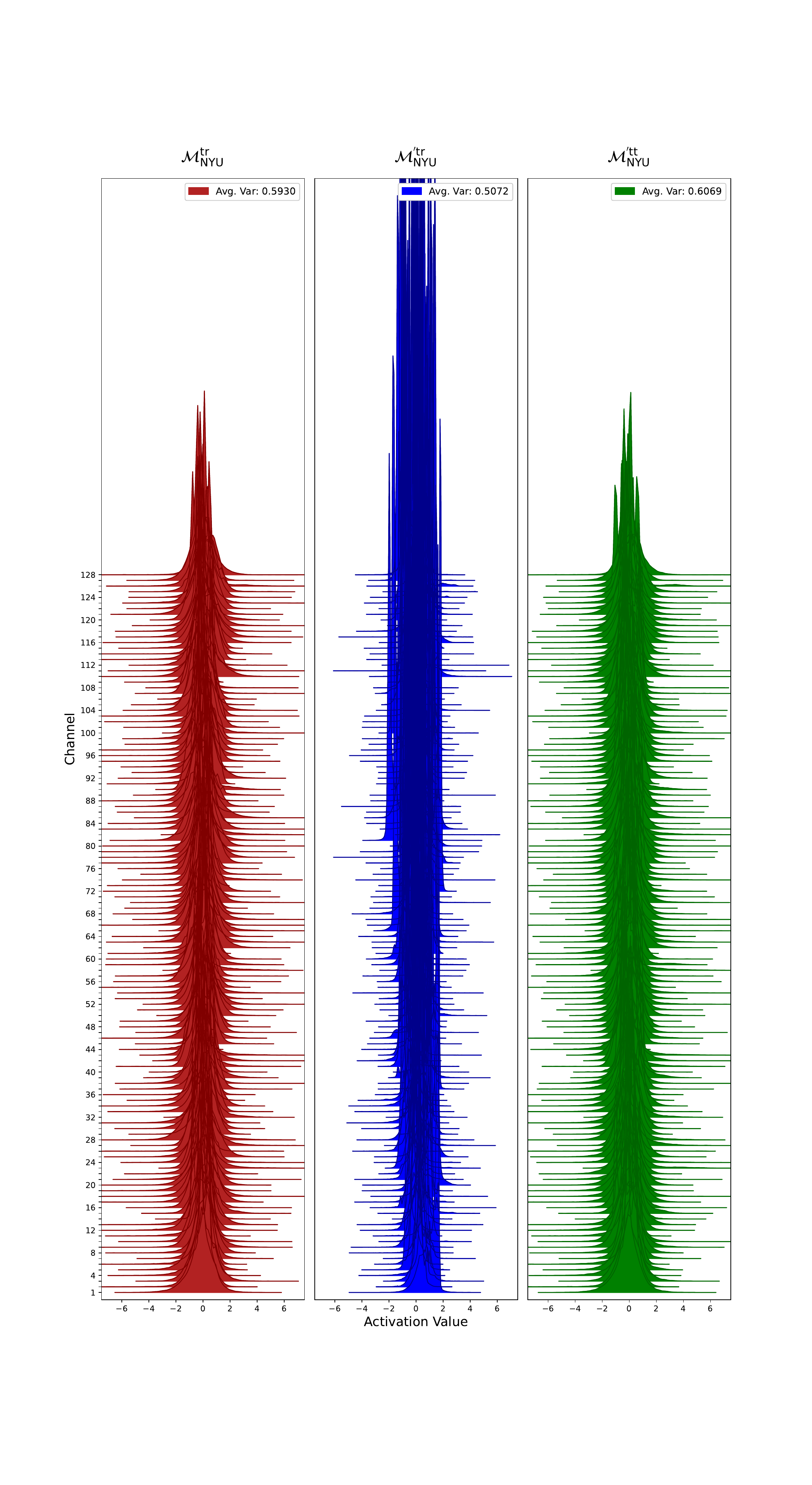}
        \vspace{-25mm}
        \caption{ResNet Layer: 4, Block: 2, BN: 2}
        \label{fig:ap:csaw:l4-b2-bn2}
    \end{subfigure}
    \vspace{-10mm}
    \caption{KDEs of per-channel activations for BN layers in the second block of ResNet layers 2, 3, and 4. All KDEs in this section are computed using a mini-batch of 16 images sampled from the CSAW dataset. The input batch was standardized for $\model{NYU}{tr}$, and normalized for both $\model{NYU}{'tr}$ and $\model{NYU}{'tt}$}
    \label{appendix:fig:bn_kde_comparison}
    \vspace*{\fill}
\end{figure}

\clearpage
\subsection{\texorpdfstring{KDEs of BN outputs from $\model{NYU}{tr}$, $\model{HCTP}{tr}$, and $\model{HCTP(BNFC)}{tr}$}{KDEs of BN outputs from NYU-tr, HCTP-tr,  and HCTP(BNFC)-tr}}

\label{appendix:kde:nyu_bnfc_hctp}
\begin{figure}[H]
    \centering
    \begin{subfigure}[t]{0.9\textwidth}
        \centering
        \includegraphics[width=\textwidth, height=5.5cm]{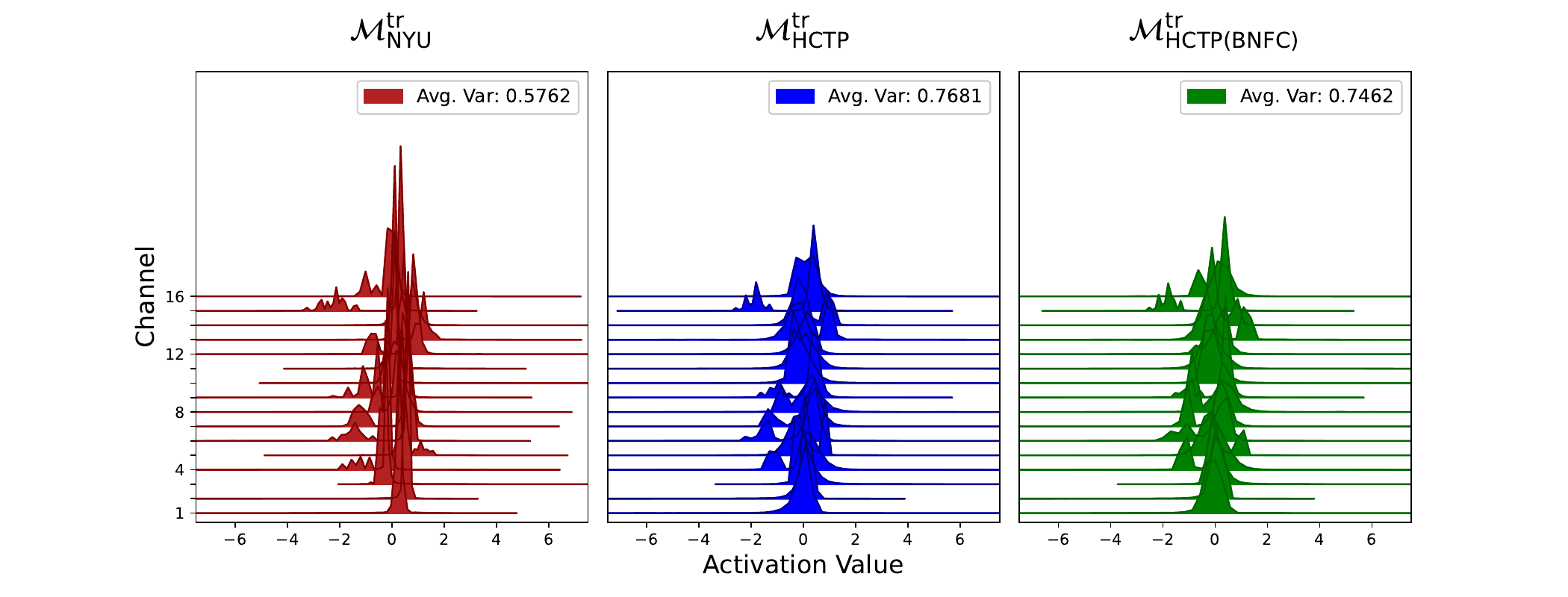}
        \caption{ResNet Layer: 1, Block: 2, BN: 2}
        \label{fig:bnfc:l1-b2-bn2}
    \end{subfigure}
    
    \begin{subfigure}[t]{0.9\textwidth}
        \centering
        \includegraphics[width=\textwidth, height=14cm]{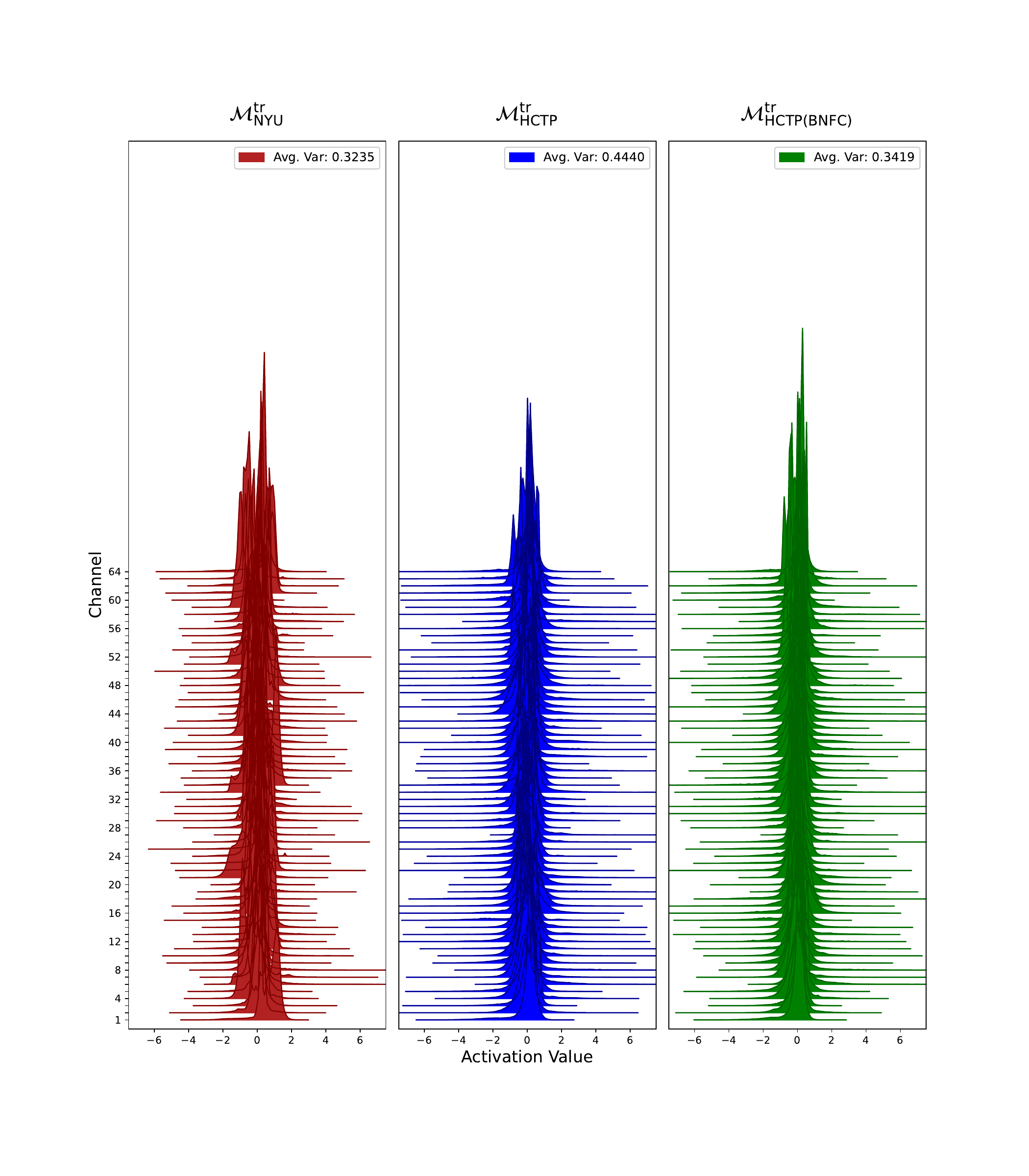}
        \vspace{-15mm}
        \caption{ResNet Layer: 3, Block: 2, BN: 2}
        \label{fig:bnfc:l3-b2-bn2}
    \end{subfigure}
    \end{figure}
    
    \begin{figure}[H]
    \ContinuedFloat
    \centering
    \begin{subfigure}[t]{0.9\textwidth}
        \centering
        \includegraphics[width=\textwidth, height=20cm]{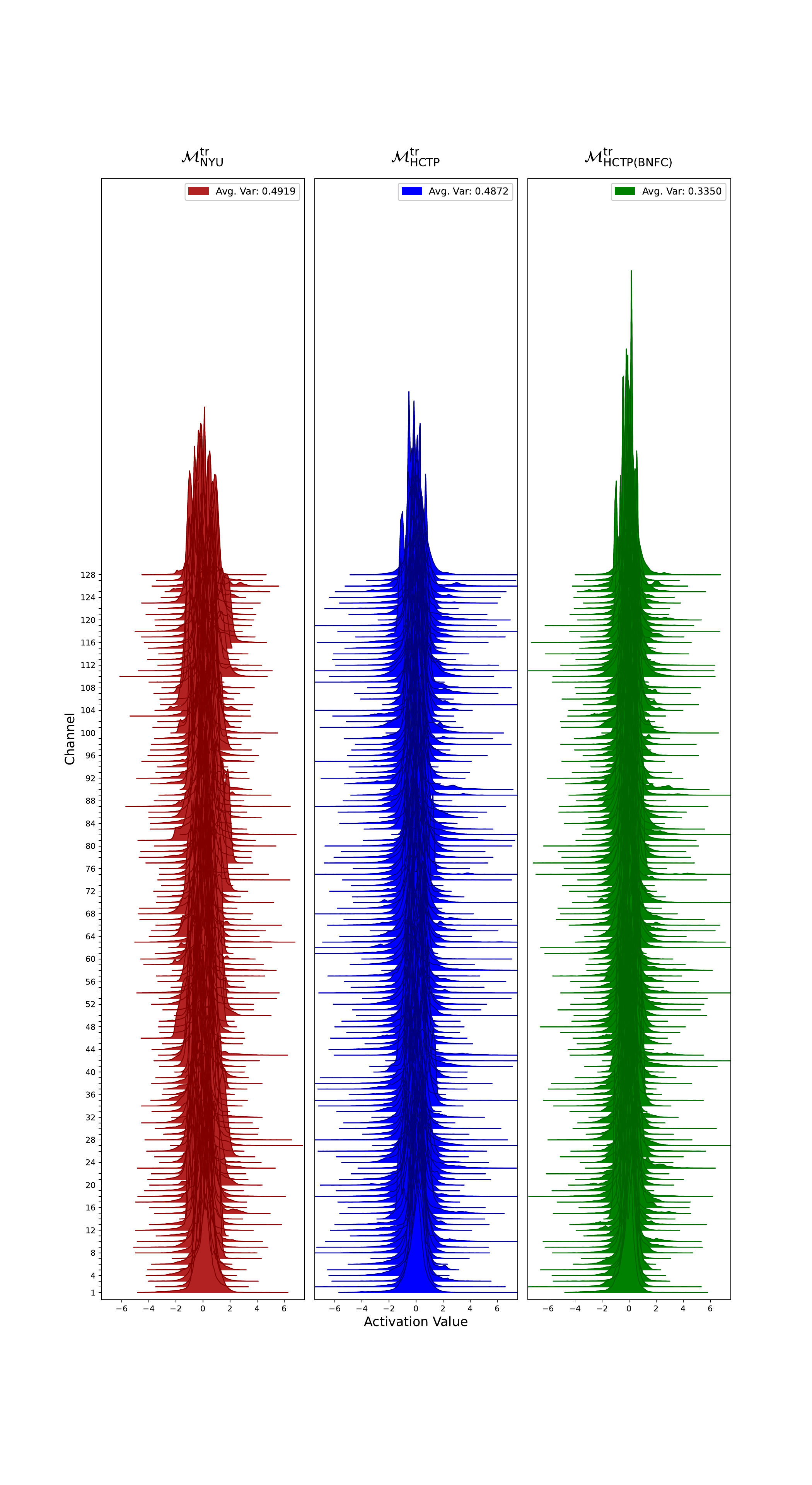}
        \vspace{-25mm}
        \caption{ResNet Layer: 4, Block: 2, BN: 2}
        \label{fig:bnfc:l4-b2-bn2}
    \end{subfigure}
    \vspace{-10mm}
    \caption{KDEs of per-channel activations for BN layers in the second block of ResNet layers 2, 3, and 4. All KDEs in this section are computed using a mini-batch of 16 images sampled from the HCTP dataset.}
    \label{appendix:fig:bn_kde_comparison:bnfc}
    \vspace*{\fill}
\end{figure}
\clearpage

\end{document}